\documentclass[twoside]{IEEEtran}
\usepackage[T1]{fontenc}
\usepackage[latin9]{inputenc}
\usepackage{float}
\usepackage{amsmath}
\usepackage{amssymb}
\usepackage{graphicx}
\usepackage{hyperref}

\hypersetup{pdftitle={Your Title},
 pdfauthor={Your Name},
 pdfpagelayout=OneColumn, pdfnewwindow=true, pdfstartview=XYZ, plainpages=false}

\makeatletter

\floatstyle{ruled}
\newfloat{algorithm}{tbp}{loa}
\providecommand{\algorithmname}{Algorithm}
\floatname{algorithm}{\protect\algorithmname}

  \newtheorem{example}{Example}
  \newtheorem{remrk}{Remark}
  \newtheorem{lemma}{Lemma}
  \newtheorem{thm}{Theorem}

\usepackage[caption=false,font=footnotesize]{subfig}
\usepackage{algorithmic}
\usepackage{cite}

\makeatother

\begin{document}

\title{Joint Frequency Reuse and Cache Optimization in Backhaul-Limited
Small-Cell Wireless Networks}

\author{\normalsize{\IEEEauthorblockN{Wei Han, An Liu, \emph{SMIEEE}, Wei
Yu, \emph{FIEEE}, and Vincent K. N. Lau, \emph{FIEEE} \thanks{This
work was supported by Science and Technology Program of Shenzhen,
China (Grant No. JCYJ20170818113908577), and RGC 16204814. The work
of An Liu was supported by the China Recruitment Program of Global
Young Experts. Wei Yu is supported by a Hong Kong Telecom Institute
of Information Technology Visiting Fellowship, and in part by an E.W.R.
Steacie Memorial Fellowship. (Corresponding author: An Liu.)}
\thanks{Wei Han is with the HKUST Shenzhen Research Institute (email: whan@connect.ust.hk).}\thanks{An
Liu is with the College of Information Science and Electronic Engineering,
Zhejiang University, Hangzhou 310027, China (e-mail: anliu@ zju.edu.cn).} \thanks{Wei
Yu is with the Electrical and Computer Engineering Department, University
of Toronto (email: weiyu@ece.utoronto.ca).}\thanks{Vincent K. N. Lau is
with the Hong Kong University of Science and Technology (email:
eeknlau@ust.hk).}}}}
\maketitle
\begin{abstract}
Caching at base stations (BSs) is a promising approach for supporting
the tremendous traffic growth of content delivery over future small-cell
wireless networks with limited backhaul. This paper considers exploiting
spatial caching diversity (i.e., caching different subsets of popular
content files at neighboring BSs) that can greatly improve the cache
hit probability, thereby leading to a better overall system performance.
A key issue in exploiting spatial caching diversity is that the cached
content may not be located at the nearest BS, which means that to
access such content, a user needs to overcome strong interference
from the nearby BSs; this significantly limits the gain of spatial
caching diversity. In this paper, we consider a joint design of frequency
reuse and caching, such that the benefit of an improved cache hit
probability induced by spatial caching diversity and the benefit of
interference coordination induced by frequency reuse can be achieved
simultaneously. We obtain a closed-form characterization of the approximate
successful transmission probability for the proposed scheme and analyze
the impact of key operating parameters on the performance. We design
a low-complexity algorithm to optimize the frequency reuse factor
and the cache storage allocation. Simulations show that the proposed
scheme achieves a higher successful transmission probability than
existing caching schemes.
\end{abstract}


\section{Introduction}

It is predicted that there will be a 1000X increase in capacity demand
for mobile data traffic in future 5G wireless networks. To meet the
rapid data traffic growth, small-cell wireless networks have been
proposed as an effective approach. By increasing the density of small-cell
base stations (BSs) deployed per unit area, the spectral efficiency
of a network can be improved. However, due to the large number of
BSs per unit area in small-cell wireless networks, allocating a high-speed
backhaul to each BS will lead to both high CAPEX and OPEX \cite{paolini2011crucial}.
In practice, the backhaul capacity of small-cell BSs is limited, and
this significantly limits the potential spectral efficiency gain provided
by small-cell networks.

Recent works show that caches can be used in small-cell wireless networks
to alleviate the high-speed backhaul capacity requirement by moving
the content closer to users \cite{wang2014cache,han2016phy,han2015degrees,liu2014cache,bacstug2015cache,yang2016analysis,liu2016cache,blaszczyszyn2015optimal,tamoor2015modeling,cui2017analysis}.
For example, in \cite{bacstug2015cache,yang2016analysis,liu2016cache,blaszczyszyn2015optimal,tamoor2015modeling,cui2017analysis},
the benefits of caching are characterized by considering the stochastic
natures of channel fading and the geographic locations of BSs and
users. The caching performance can be analyzed and optimized using
the theory of stochastic geometry. Specifically, in \cite{bacstug2015cache},
the authors consider a cache placement scheme in which all BSs store
the same set of the most popular content files, and then analyze the
outage probability and average rate. The uncached files are served
using the data obtained from the backhaul, so the service rates are
limited by the backhaul capacity. Likewise, the authors of \cite{yang2016analysis}
analyze the average ergodic rate and the outage probability in a three-tier
heterogeneous network with backhaul capacity constraints and with
the caching of the most popular files. Caching the same subset of
the most popular files at every BS is, however, not optimal in general.
In \cite{blaszczyszyn2015optimal}, the authors consider random caching
at BSs and analyze the cache hit probability under general popularity
distribution (but without considering backhaul constraints), and show
that it is not always optimal to cache the most popular content files
in every BS. The reason behind this is that placing different contents
in different BSs provides \emph{spatial caching diversity}, which
brings better overall performance. 

\begin{figure}[t]
\begin{centering}
\includegraphics[width=1\columnwidth]{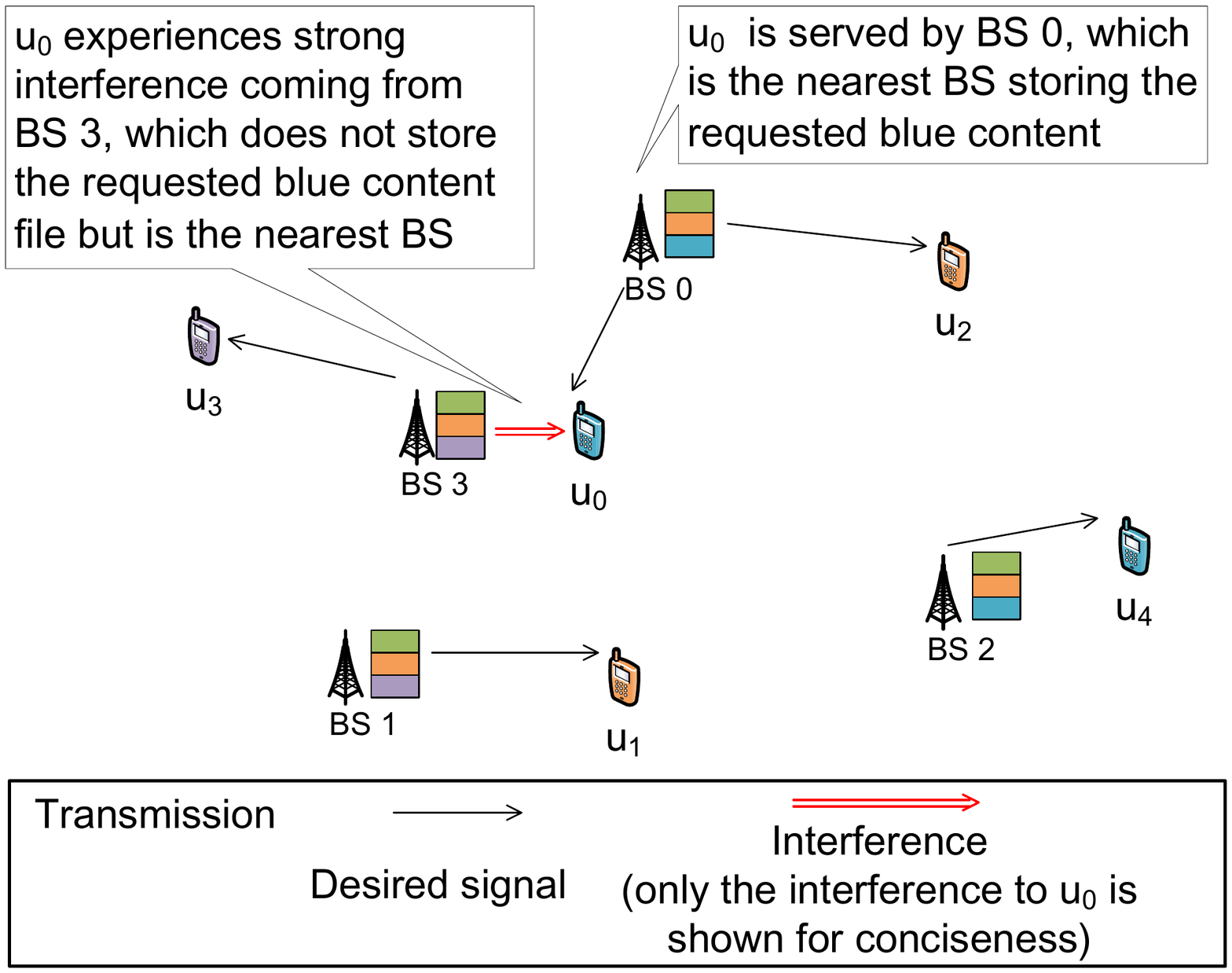}
\par\end{centering}
\caption{\label{fig:single-band-network}Illustration of the strong interference
in single-band random caching schemes. $u_{0}$ is served by BS 0,
which has its requested blue content file, but is not the geographically
nearest BS of $u_{0}$. In this case, $u_{0}$ experiences strong
interference coming from BS 3, which is the geographically nearest
BS of $u_{0}$.}
\end{figure}

Spatial caching diversity is typically achieved by random caching
strategies in the existing literature. For example, in \cite{tamoor2015modeling},
the authors study caching in a wireless network with uniform content
popularity distribution but without a backhaul constraint, and analyze
cache hit probability and content outage probability for random caching
with a uniform distribution. In \cite{cui2017analysis}, the authors
consider a heterogeneous wireless network which caches the same subset
of the most popular files at the macro-BSs but uses random caching
at the pico-BSs; they analyze and optimize the successful transmission
probability in the high signal-to-noise ratio (SNR) and user density
regions with a backhaul capacity constraint, where the uncached files
are served by macro BSs using the data obtained from the backhaul.
Note that in \cite{blaszczyszyn2015optimal,tamoor2015modeling,cui2017analysis},
spatial caching diversity is achieved by randomly caching different
files at different BSs, as illustrated in Fig. \ref{fig:single-band-network}.
In such cases, a user may be served by a BS which has its requested
content file but is not the geographically nearest BS. This may result
in strong interference coming from the geographically nearest BS for
the target user. Hence, the benefit of spatial caching diversity may
be overwhelmed by excessive inter-cell interference.

Spatial caching diversity can also be achieved with coded caching
by encoding each content file into coded bits and caching different
portions of the coded bits in different BSs \cite{bioglio2015optimizing,liao2017optimizing,xu2017modeling}.
Specifically, in \cite{bioglio2015optimizing} and \cite{liao2017optimizing},
a maximum distance separable (MDS)-coded caching scheme is considered.
However, the physical layer is modeled by error-free links between
users and their associated BSs, and the effect of interference is
ignored. In \cite{xu2017modeling}, multiple BSs cache different coded
packets of each file, and each user employs successive interference
cancellation (SIC) to cancel the strong interference from the nearest
BSs before decoding the desired signal. Although SIC removes strong
interference from the nearest BSs, the complexity of the receiver
at the user side increases. Moreover, the analysis in \cite{xu2017modeling}
is obtained under the simplified assumption that each BS transmits
signals all the time, and only one typical user is considered. Also,
the resource allocation for multi-user transmission at each BS and
the effect of system loading are not studied in \cite{xu2017modeling}. 

This paper proposes to address the interference induced by spatial
caching diversity by joint design of frequency reuse and caching.
Frequency reuse is a well-studied inter-cell interference coordination
technique in conventional link-based wireless networks \cite{katzela1996channel}.
In a system with frequency reuse, two adjacent cells may use different
frequencies to reduce the strong interference experienced by the cell
edge users. In this way, both coverage and capacity are improved \cite{miao2016fundamentals}. 

In this paper, we propose to explore the joint design of frequency
reuse and caching, such that the benefit of spatial caching diversity
and frequency reuse can be achieved at the same time. We consider
a content delivery application with a fixed data rate requirement
for each user. In such an application, the performance is characterized
by the successful transmission probability. The main contributions
of this paper are summarized as follows:
\begin{itemize}
\item \textbf{A joint design of frequency reuse and caching scheme:} In
this paper, we propose a joint frequency reuse and caching scheme
such that the subset of BSs allocated the same frequency also caches
the same subset of content files. By such joint design, the strong
interference caused by spatial caching diversity can be removed and
a higher successful transmission probability can be achieved.
\item \textbf{Closed-form characterization of the approximate successful
transmission probability:} To analyze the impact of key operating
parameters (such as number of subbands and cache storage capacity
allocation) on the system performance, we derive a closed-form characterization
of the approximate successful transmission probability under the joint
frequency reuse and caching scheme.
\item \textbf{Optimization of the frequency reuse factor and cache placement:}
The problem of optimizing the number of subbands and the cache storage
capacity allocation is a complex integer optimization problem. We
propose a low-complexity algorithm and show that the proposed scheme
achieves a large gain over existing caching schemes in terms of the
successful transmission probability.
\end{itemize}

The rest of the paper is organized as follows. The model of caching
in backhaul-limited small-cell wireless network under study is presented
in Section \ref{sec:System-Model}. A joint frequency reuse and caching
scheme is proposed in Section \ref{sec:Joint-Frequency-Reuse}. In
Section \ref{sec:Performance-Metric-and}, we define the average successful
transmission probability as the performance metric, and analyze the
performance of the proposed scheme for a given cache storage allocation
and frequency reuse factor. In Section \ref{sec:Optimization-of-Cache},
we formulate and solve the joint cache storage allocation and frequency
reuse optimization problem. Numerical evaluation of the proposed scheme
is presented in Section \ref{sec:Simulation-Results}. We conclude
the paper in Section \ref{sec:Conclusion}.

\section{System Model\label{sec:System-Model}}

We consider a backhaul-limited small-cell wireless network. The locations
of the BSs are spatially distributed as a homogeneous Poisson point
process (PPP) $\Phi^{b}$ with density $\lambda^{b}$. The locations
of the users are also spatially distributed as a homogeneous PPP $\Phi^{u}$
with density $\lambda^{u}$. There is a content library $\mathcal{X}=\left\{ X_{1},X_{2},\dots,X_{L}\right\} $
that contains $L$ files, where the size of each content file is $F$
bits. Each content file is independently requested with probability
$\rho_{l}$, satisfying $\sum_{l=1}^{L}\rho_{l}=1$. Without loss
of generality, we assume $\rho_{1}\geq\rho_{2}\geq\dots\geq\rho_{L}$. 

We consider the downlink transmission, where the content file requested
by each user is transmitted at a fixed rate of $\tau$ bits per second.
Each BS has one transmit antenna with transmission power $P$, a cache
of storage capacity $B_{C}F$ bits, and a backhaul with limited capacity
of $B_{B}\tau$ bits per second (bps). Each user has one receive antenna.
The total bandwidth is $W$ Hz. We consider a discrete-time system,
with time being slotted with duration $\nu$, and study one time slot
of the network. We consider both large-scale fading (path loss) and
small-scaling fading. Specifically, the channel coefficient between
a BS and a user with distance $D$ is modeled by $D^{-\alpha}h$,
where $\alpha>2$ is the path loss exponent, $h\overset{d}{\sim}\mathcal{CN}\left(0,1\right)$
is the small scale fading factor (i.e., we assume Rayleigh fading
channels).

\section{Joint Frequency Reuse and Caching Scheme\label{sec:Joint-Frequency-Reuse}}

In this section, we propose a \emph{joint frequency reuse and caching
scheme} which exploits the benefit of interference coordination induced
by frequency reuse and the benefit of backhaul offloading induced
by spatial caching diversity. 

\subsection{Joint Frequency Reuse and Cache Placement}

The BSs are randomly divided into $M$ non-overlapping BS groups indexed
by $\left\{ 0,\dots,M-1\right\} $. Specifically, each BS independently
and randomly generates a number from $\left\{ 0,\dots,M-1\right\} $,
say $m$, and then joins the $m$-th BS group. Denote $\Phi_{m}^{b}$
with $m\in\left\{ 0,\dots,M-1\right\} $ as the BSs in the $m$-th
BS group. For analysis purposes, we assume random BS grouping and
independent thinning \cite[p. 230]{haenggi2009interference} so that
$\Phi_{m}^{b}$ follows a homogeneous PPP with density $\frac{\lambda_{b}}{M}$.
The total bandwidth $W$ is also divided into $M$ equal-size subbands
denoted as $W_{0},W_{1}\dots,W_{M-1}$, where the bandwidth of each
subband is $\frac{W}{M}$. The BSs in $\Phi_{m}^{b}$ are designed
to transmit in subband $W_{m}$ for $m=0,\dots,M-1$. 

\begin{figure}[t]
\begin{centering}
\includegraphics[width=1\columnwidth]{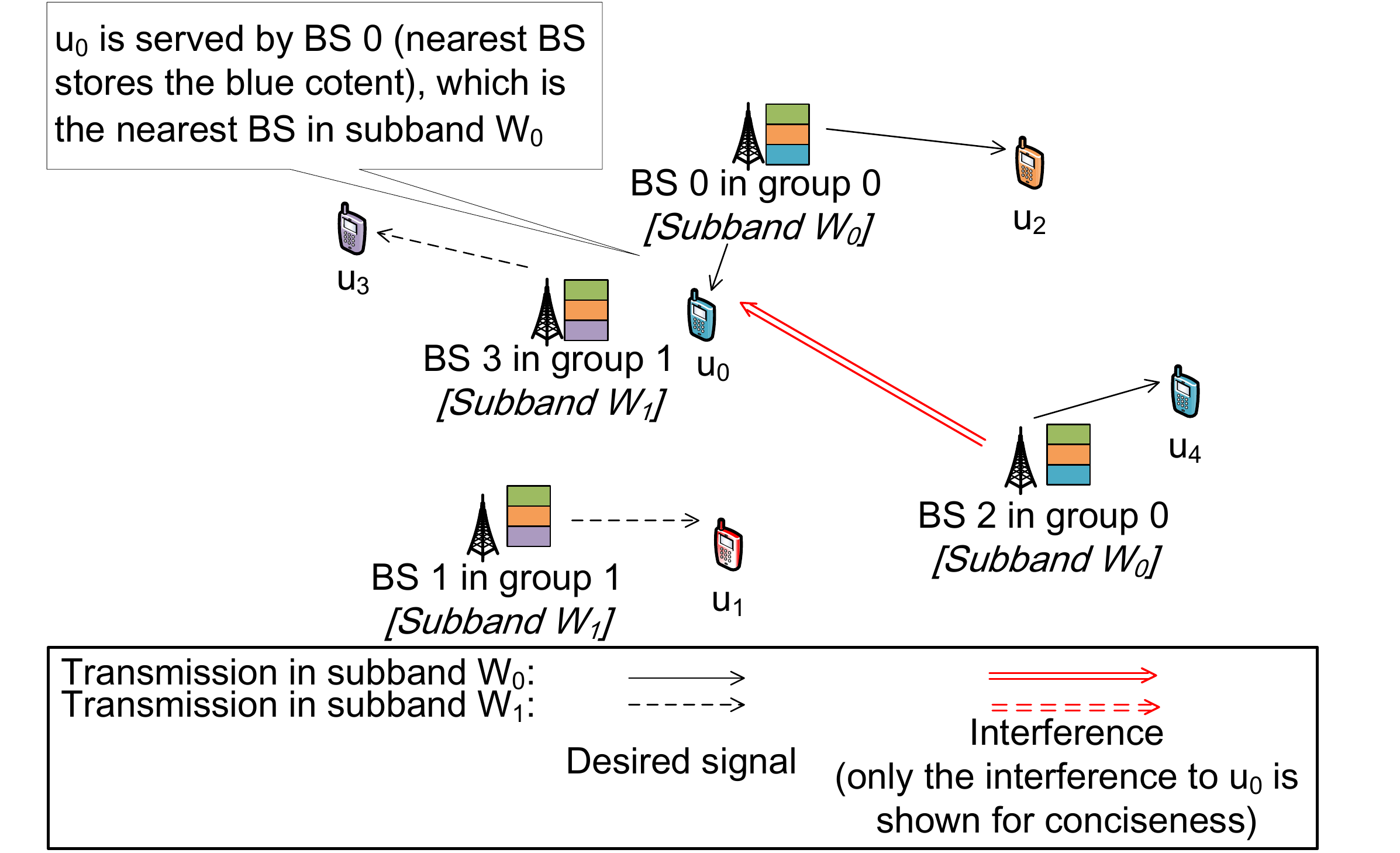}
\par\end{centering}
\caption{\label{fig:Network-model}Illustration of the joint frequency reuse
and cache placement scheme. The user $u_{0}$ is served by BS 0, which
has $u_{0}$'s requested blue content file, and BS 0 is the geographically
nearest BS in the transmitting subband $W_{0}$. The geographically
nearest BS (BS 3) is transmitting in subband $W_{1}$, and does not
cause interference to $u_{0}$.}
\end{figure}

The proposed joint frequency reuse and cache placement design helps
mitigate inter-cell interference in the network. Note that in \cite{blaszczyszyn2015optimal},
the authors propose a probabilistic cache placement policy, which
sets the probability of storing each content at a given BS to an optimized
target value. Such a design does not consider the strong interference
induced by spatial caching diversity, because in the scheme proposed
in \cite{blaszczyszyn2015optimal}, a user may be served by a BS that
has its requested content file but is not the geographically nearest
BS, which may result in the user experiencing strong interference
from the geographically nearest BS, as illustrated in Fig. \ref{fig:single-band-network}.
In this paper, we propose a joint frequency reuse and cache placement
strategy that can mitigate inter-cell interference in the network.
In our scheme, the BSs in one BS group store the same subset of content
files. As a user is served by the nearest BS that stores the requested
content file, the serving BS must also be the geographically nearest
BS in its transmitting subband (BS group), which leads to a higher
receiving signal-to-interference-plus-noise ratio (SINR) at the user,
as illustrated in Fig. \ref{fig:Network-model}.

\begin{figure}[t]
\begin{centering}
\includegraphics[width=1\columnwidth]{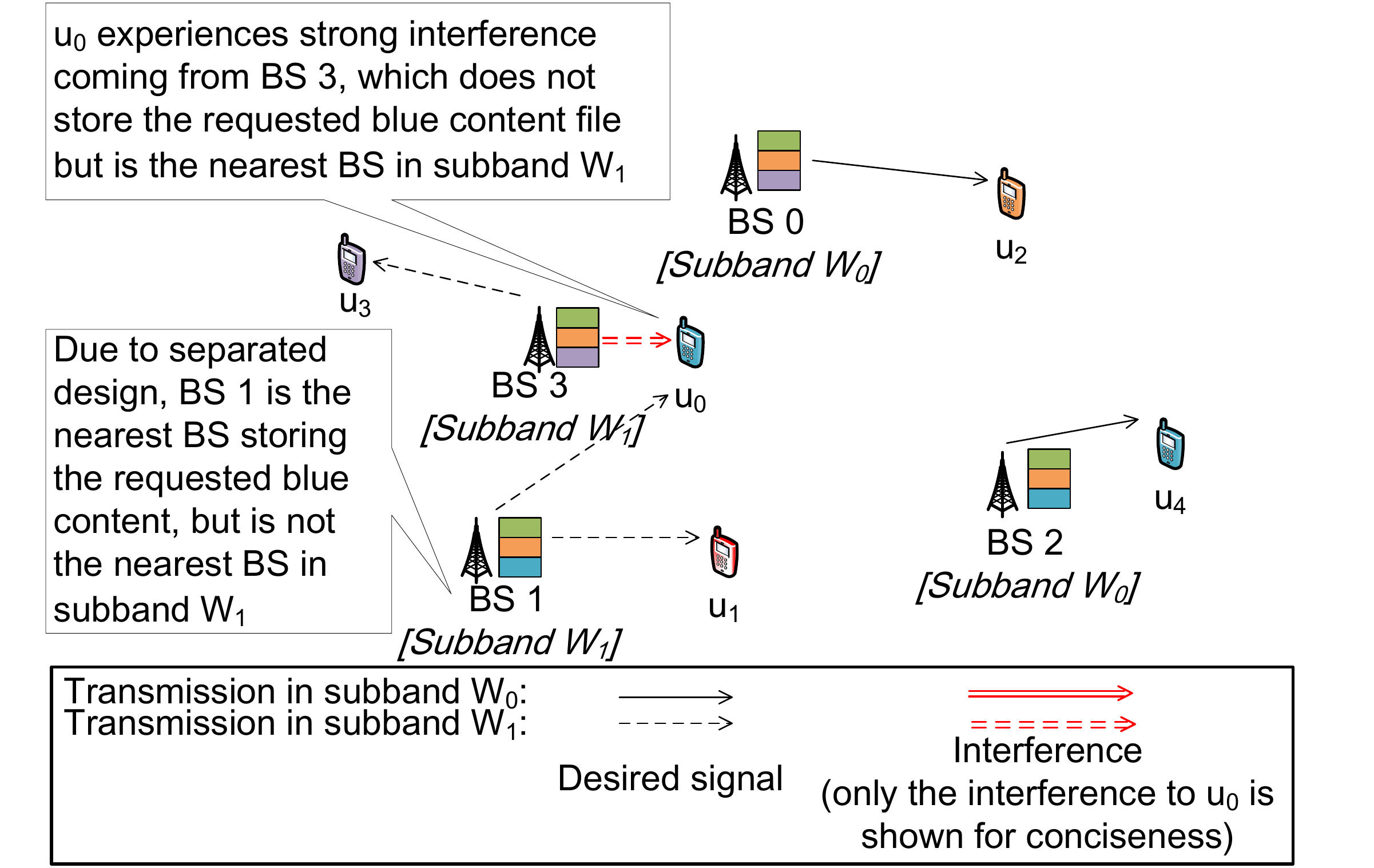}
\par\end{centering}
\caption{\label{fig:separate-network}Illustration of the strong interference
caused by separated design of frequency reuse and cache placement.
The frequency reuse factor is $\frac{1}{2}$, and the cache placement
follows random caching. Due to separated design, $u_{0}$ is served
by BS 1, which has its requested blue content file, but is not the
geographically nearest BS in the transmitting subband $W_{1}$. In
this case, $u_{0}$ experiences strong interference coming from BS
3, which is the geographically nearest BS in the transmitting subband
$W_{1}$.}
\end{figure}

Note that a naive combination of the cache placement scheme in \cite{blaszczyszyn2015optimal}
and frequency reuse cannot completely address the strong interference
issue. If the frequency reuse and cache placement are designed separately,
the nearest BS storing the requested content file would not necessarily
be the geographically nearest BS in its transmitting subband. In this
case, the user may experience strong interference coming from the
geographically nearest BS in the transmitting subband, as illustrated
in Fig. \ref{fig:separate-network}. Numerical results in Section
\ref{sec:Simulation-Results} also show that our proposed joint scheme
outperforms naive combination of the cache placement scheme in \cite{blaszczyszyn2015optimal}
and frequency reuse.

A main contribution of this paper is to design the optimal caching
policy of the content files. Each content file may be stored in multiple
BS groups. Denote $q_{l}\in\left\{ 0,1,\dots,M\right\} $ for $l\in\left\{ 1,\dots,L\right\} $
as the \emph{cache storage allocation factor}, which indicates that
the $l$-th content file is stored in a total of $q_{l}$ BS groups.
We assume that $q_{l}\geq q_{l+1}$, for $l=1,\dots,L-1$, i.e., a
content file with higher popularity is stored in more BSs. Denote
$\mathbf{q}=\left[q_{1},\dots,q_{L}\right]$ as the \emph{cache storage
allocation vector}, which must satisfy the following cache storage
capacity constraint:
\begin{equation}
\sum_{l=1}^{L}q_{l}\leq MB_{C}.\label{eq:cache-constraint}
\end{equation}

\begin{figure}[t]
\begin{centering}
\includegraphics[width=1\columnwidth]{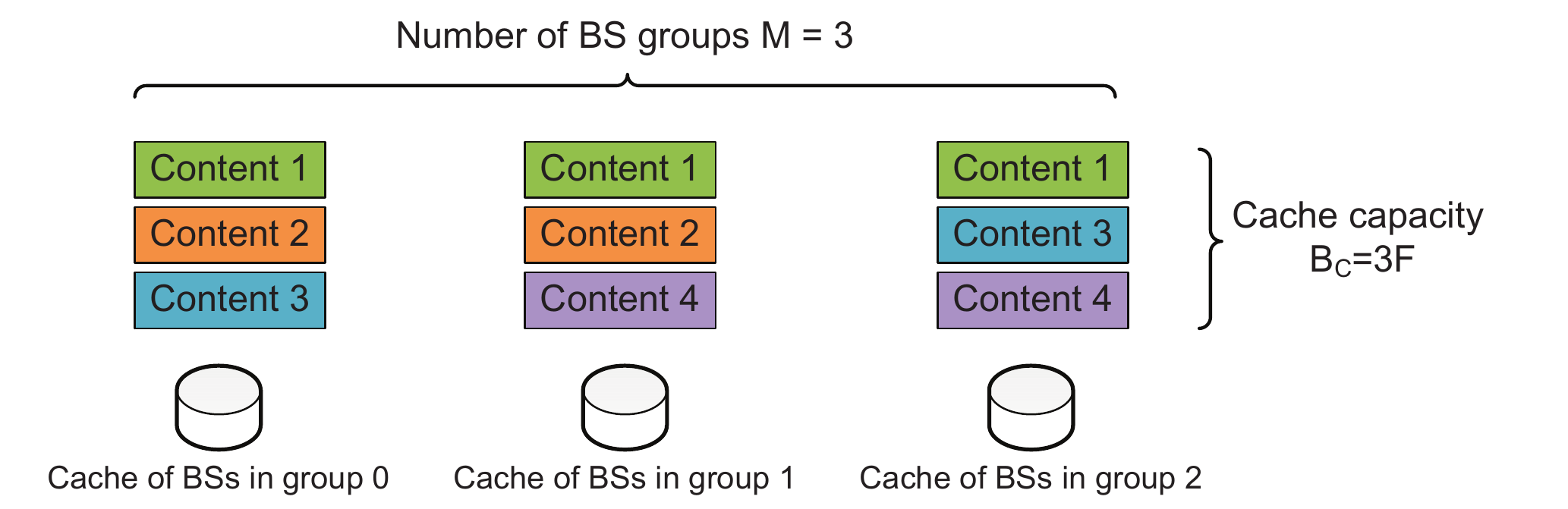}
\par\end{centering}
\caption{\label{fig:Illustration-of-cache}Illustration of the cache placement
and cache data structure in Example \ref{example:Caching-scheme}.}
\end{figure}

The proposed cache placement is illustrated in Fig. \ref{fig:Illustration-of-cache}
for a network with $M=3$ BS groups and cache storage capacity $B_{C}=3$
files at each BS. The detailed cache data structure is elaborated
below. Since all BSs in the same BS group cache the same subset of
content files, we can use a single cache memory with $B_{C}$ memory
blocks to represent the cache data structure for each BS group. First,
the cache memory for each BS group is divided into $B_{C}$ memory
blocks of size $F$ bits and each memory block caches one content
file. The cached content for all BS groups can be arranged in a matrix
form, where the $\left(n,m\right)$-th entry is the $n$-th memory
block for the $m$-th BS group, as illustrated in Fig. \ref{fig:Illustration-of-cache}.
For a given cache storage allocation vector $\mathbf{q}$ satisfying
(\ref{eq:cache-constraint}), the $L$ content files are placed one
after another to fill in the cache memory sequentially from left to
right and top to bottom in the matrix of cache memory blocks, where
the $l$-th content file fills a total number of $q_{l}$ cache memory
blocks. Specifically, if $q_{l}=0$, then the $l$-th content file
is not stored in any of the BS caches, and will be served via the
backhaul at the BSs. If $q_{l}\neq0$, then the $l$-th content file
is stored in the cache of the BSs in $\bigcup_{m\in\mathcal{M}_{l}}\Phi_{m}^{b}$,
where $\mathcal{M}_{l}$ is the index of the BS groups that store
the $l$-th content file, given by
\begin{align}
\mathcal{M}_{l}= & \bigg\{\mathrm{mod}\left(\sum_{l^{\prime}=1}^{l-1}q_{l^{\prime}},M\right),\mathrm{mod}\left(\sum_{l^{\prime}=1}^{l-1}q_{l^{\prime}}+1,M\right),\nonumber \\
 & \dots,\mathrm{mod}\left(\sum_{l^{\prime}=1}^{l-1}q_{l^{\prime}}+q_{l}-1,M\right)\bigg\}.
\end{align}
The proposed scheme ensures that for a given $\mathbf{q}$ satisfying
(\ref{eq:cache-constraint}), the number of content files stored in
each cache is the same, and each BS caches $B_{C}$ distinct content
files. The proposed cache placement scheme is illustrated in the following
example.

\begin{example}
[Cache placement] \label{example:Caching-scheme}Consider a network
with $M=3$ BS groups, cache size $B_{C}F=3F$, and total number of
content files $L=6$. The cache storage allocation vector is given
by $\mathbf{q}=\left[3,2,2,2,0,0\right]$, satisfying the cache storage
capacity constraint given in (\ref{eq:cache-constraint}). As illustrated
in Fig. \ref{fig:Illustration-of-cache}, the first content file is
stored in three BS groups indexed by $\mathcal{M}_{1}=\left\{ 0,1,2\right\} $.
Each of the other content files is stored in two BS groups, given
by $\mathcal{M}_{2}=\left\{ 0,1\right\} $, $\mathcal{M}_{3}=\left\{ 0,2\right\} $,
$\mathcal{M}_{4}=\left\{ 1,2\right\} $, and $\mathcal{M}_{5}=\mathcal{M}_{6}=\emptyset$.
It can be easily seen that each BS caches three distinct content files,
which satisfies the cache storage capacity constraint.
\end{example}

Note that for practical consideration, the initialized cache content
at the BSs does not adapt to the instantaneous realization of the
user request at fast timescale. Instead, $\mathbf{q}$ is adaptive
only to the content popularity statistics. As a result, the BS cache
update is done over a slow timescale when the network is lightly loaded.

\subsection{Content Delivery\label{subsec:Transmission-Scheme}}

\subsubsection{Content-Centric User Scheduling}

We adopt a content-centric user scheduling scheme. Different from
the conventional connection-based user scheduling scheme, which is
based on physical layer parameters, this content-centric user scheduling
scheme jointly considers both the physical layer and content status
of BS caches. Consider a user which requests the $l$-th content file.
If the requested $l$-th content file is stored in some of the BS
caches in the network, i.e., $q_{l}\neq0$, then the user is associated
with the nearest BS which stores the $l$-th content file in its cache.
Otherwise, if the requested $l$-th content file is not stored in
any BS caches, i.e., $q_{l}=0$, then the user is associated with
the geographically nearest BS in $\Phi^{b}$ (and the content file
is fetched via the backhaul). The proposed user association scheme
is illustrated using the example in Fig. \ref{fig:Network-model}.
The blue content file requested by $u_{0}$ is stored in the cache
of BS 0 and BS 2; in this case $u_{0}$ is associated with BS 0, which
is the nearest BS which stores the blue content file. The red content
file requested by $u_{1}$ is not stored in any BS caches, then $u_{1}$
is associated with BS 1, which is the geographically nearest BS, and
the red content file is fetched via the backhaul. A similar content-centric
user association has also been adopted in existing works on cached
wireless networks \cite{blaszczyszyn2015optimal} and \cite{cui2017analysis}.
Compared with the conventional nearest BS association, the content-centric
user association schemes in our paper, \cite{blaszczyszyn2015optimal}
and \cite{cui2017analysis} require some additional information about
user requests and content placement at the BSs. Since the user requests
change at a much longer timescale compared with the duration of a
time slot, the induced additional overhead is low, and is practically
feasible.

Without loss of generality, we study the performance of a typical
user, which is located at the origin. Denote $u_{0}$ as the typical
user, and denote $B_{0}$ as the serving BS of $u_{0}$. Denote $K_{l}$
as the number of users associated with $B_{0}$ which request the
$l$-th content file, and denote $\mathbf{K}=\left[K_{1},\dots,K_{L}\right]$
as the BS loading vector. $\mathbf{K}$ may take value $\mathbf{k}=\left[k_{1},\dots,k_{L}\right]$,
where $k_{l}\in\left\{ 0,1,\dots\right\} $. Note that not all associated
users of $B_{0}$ can always be served by $B_{0}$ at the same time.
Due to the limited backhaul transmission capacity, if more than $B_{B}$
uncached content files are requested from one BS, then $B_{B}$ users
are randomly selected to be served with the content files obtained
from the backhaul.\footnote{Note that even when two users request the same content file from the
backhaul, the probability that the two users request the same portion
of the content file within the current time slot is very small for
the typical content file size and slot duration, and thus they still
need to consume a backhaul capacity of 2$\tau$ bits/s.} Denote $S$ as the user scheduling state, where $S=1$ represents
the event that $u_{0}$ is scheduled to be served by $B_{0}$, and
$S=0$ represents the event that $u_{0}$ is not scheduled to be served.
The user scheduling is illustrated in the following example.
\begin{example}
[User scheduling] \label{example:backhaul-outage}Consider a network
with $M=3$ BS groups, backhaul transmission capacity $B_{B}\tau=2\tau$,
cache capacity $B_{C}F=3F$ at each BS, and the total number of content
files $L=6$. The cache storage allocation vector is given by $\mathbf{q}=\left[3,2,2,2,0,0\right]$,
and the BS loading vector is given by $\left[5,4,4,3,3,2\right]$,
which indicates that among the users associated with $B_{0}$, the
number of users requesting the first content file is 5, the number
of users requesting the second content file is 4, and so on. Since
$q_{l}\neq0,\ l\in\left\{ 1,\dots,4\right\} $, the users requesting
the first four content files can be served using cached data. However,
since $q_{5}=q_{6}=0$, users requesting the fifth and sixth content
files are served via the data obtained from the backhaul. Due to the
backhaul transmission capacity constraint, two users will be randomly
selected from the five users requesting the fifth and sixth content
files to be served using the date obtained from the backhaul. 
\end{example}

\subsubsection{PHY Transmission}

We adopt unicast and frequency division multiple access (FDMA) with
uniform bandwidth and transmit power allocation for the users associated
with each BS.\footnote{In practice, each file consists of a large number of segments, and
the probability of two users requesting the same segment at the same
time is small. As pointed out in \cite{ji2015throughput}, even though
users keep requesting the same few popular files, the asynchronism
of their requests is usually large with respect to the duration of
the file (e.g., video) itself, such that the probability that a single
transmission from the source nodes is useful for more than one user
(i.e., multicasting) is essentially zero. This phenomenon is called
``\textit{asynchronous content reuse}'' in \cite{ji2015throughput}.
As a result, ``naive'' multicasting due to repeated requests for the
same segment of the same file at the same time from different users
is unlikely to occur in practice.} Consider one BS which simultaneously transmits to a total number
of $G_{0}$ associated users. The BS transmits each of the associated
users at a rate of $\tau$ bps over bandwidth $\frac{W}{MG_{0}}$.
The transmit power is proportional to the allocated bandwidth, given
by $\frac{P}{G_{0}}$. We assume that all BSs are active. When $u_{0}$
is served with file $l_{0}$, the received signal of $u_{0}$ is given
by
\begin{equation}
y_{0}=D_{0,0}^{-\frac{\alpha}{2}}h_{0,0}x_{0}+\sum_{n\in\Phi_{m_{0}}^{b}\backslash B_{0}}D_{n,0}^{-\frac{\alpha}{2}}h_{n,0}x_{n}+z_{0},
\end{equation}
where $D_{0,0}$ is the distance between $B_{0}$ and $u_{0}$, $h_{0,0}\sim\mathcal{CN}\left(0,1\right)$
is the small-scale channel fading between $B_{0}$ and $u_{0}$, $x_{0}$
is the transmit signal from $B_{0}$ to $u_{0}$ satisfying the transmit
power constraint $\mathbb{E}\left(\left\Vert x_{0}\right\Vert \right)=\frac{P}{G_{0}}$,
$\Phi_{m_{0}}^{b}$ is the group of BSs which $B_{0}$ belongs to
(i.e., $B_{0}\in\Phi_{m_{0}}^{b}$), $D_{n,0}$ is the distance between
BS $n$ and $u_{0}$, $h_{n,0}\sim\mathcal{CN}\left(0,1\right)$ is
the small-scale channel fading between BS $n$ and the typical user
$u_{0}$, $x_{n}$ is the transmit signal from BS $n$ to its associated
user in the $m_{0}$-th frequency band satisfying transmit power constraint
$\mathbb{E}\left(\left\Vert x_{n}\right\Vert \right)=\frac{P}{G_{0}}$,
$z_{0}$ is the complex additive white Gaussian noise of power $\frac{WN_{0}}{MK_{0}}$,
and $N_{0}$ is the noise spectral density. In this paper, we consider
the high SINR regime where $P/W\gg N_{0}$. The signal-to-interference
ratio (SIR) of $u_{0}$ is given by
\begin{equation}
\mathrm{SIR}=\frac{D_{0.0}^{-\alpha}\left|h_{0,0}\right|^{2}}{\sum_{n\in\Phi_{m_{0}}^{b}\backslash B_{0}}D_{n.0}^{-\alpha}\left|h_{n,0}\right|^{2}}.
\end{equation}
In the interference-limited regime, the achievable rate of $u_{0}$
is given by
\begin{equation}
C=\frac{W}{MG_{0}}\log_{2}\left(1+\mathrm{SIR}\right).
\end{equation}

\section{Performance Metric and Analysis\label{sec:Performance-Metric-and}}

The requested content file can be decoded correctly at $u_{0}$ only
when $u_{0}$ is scheduled to be served by $B_{0}$ (i.e., $S=1$)
and the physical layer achievable rate is larger than the target rate
(i.e., $C\geq\tau$). Therefore, the average successful transmission
probability is defined as
\begin{align}
 & p\left(M,\mathbf{q}\right)\nonumber \\
 & \triangleq\Pr\left[C\geq\tau,\ S=1\right]\\
 & =\Pr\left[S=1\right]\Pr\left[C\geq\tau\big|S=1\right]\\
 & =\mathbb{E}_{\mathbf{K},L_{0}}\Pr\left[S=1\big|\mathbf{K},L_{0}\right]\Pr\left[C\geq\tau\big|\mathbf{K},L_{0},S=1\right],
\end{align}
where $L_{0}$ is the random user request from the typical user $u_{0}$,
and $L_{0}=l_{0}$ represents that the $l_{0}$-th content file is
requested by $u_{0}$. The probability is with respect to the distribution
of the random user requests $L_{0}$, index of the BS group $m_{0}$
that $B_{0}$ belongs to, BS loading $\mathbf{K}$, large-scale channel
fading $D_{0.0}^{-\frac{\alpha}{2}}$, and small-scale channel fading
$h_{0,0}$. Note that the number of BS groups and subbands $M$ and
cache storage capacity allocation vector $\mathbf{q}$ fundamentally
determine the average successful transmission probability that can
be achieved. As a result, we explicitly write $p\left(M,\mathbf{q}\right)$
as a function of $M$ and $\mathbf{q}$. We will analyze the \textit{conditional
user scheduling probability} $\Pr\left[S=1\big|\mathbf{K}=\mathbf{k},L_{0}=l_{0}\right]$
and \textit{conditional physical layer successful transmission probability}
$\Pr\left[C\geq\tau\big|\mathbf{K}=\mathbf{k},L_{0}=l_{0},S=1\right]$,
respectively, in the following two subsections. 

\subsection{Conditional User Scheduling Probability}

Under the proposed content-centric user scheduling scheme, the number
of users associated with $B_{0}$ whose requested content files exist
in the cache is $\sum_{l\in\left\{ l\big|q_{l}\neq0\right\} }K_{l}$,
and the number of users associated with $B_{0}$ whose requested content
files do not exist in the cache and have to be fetched from the backhaul
is $\sum_{l\in\left\{ l\big|q_{l}=0\right\} }K_{l}$. Therefore, the
total number of users simultaneously served by $B_{0}$ is given by
\begin{equation}
G_{0}=\sum_{l\in\left\{ l\big|q_{l}\neq0\right\} }K_{l}+\min\left\{ \sum_{l\in\left\{ l\big|q_{l}=0\right\} }K_{l},B_{B}\right\} ,\label{eq:loading}
\end{equation}
where the minimum in the second term is due to the limited backhaul
transmission capacity. In general, we have $G_{0}\leq\sum_{l=1}^{L}K_{l}$,
and if $G_{0}<\sum_{l=1}^{L}K_{l}$, the $\sum_{l=1}^{L}K_{l}-G_{0}$
unserved users will suffer from outage caused by the limited backhaul
capacity. As a result, the conditional user scheduling probability
is given by
\begin{align}
 & \Pr\left[S=1\big|\mathbf{K}=\mathbf{k},L_{0}=l_{0}\right]\nonumber \\
 & =\begin{cases}
1, & q_{l_{0}}\neq0,\\
\min\left\{ \frac{B_{B}}{\sum_{l\in\left\{ l\big|q_{l}=0\right\} }k_{l}},1\right\} , & q_{l_{0}}=0.
\end{cases}\label{eq:backhaul-outage}
\end{align}

\begin{remrk}
Equation (\ref{eq:backhaul-outage}) illustrates the following effect
of the caching allocation strategy \textbf{$\mathbf{q}$} and the
number of BS groups (i.e. subbands) $M$ on the conditional user scheduling
probability:
\begin{enumerate}
\item \textbf{Effect of $\mathbf{q}$:}
\begin{enumerate}
\item $q_{l_{0}}\neq0$: The requested $l_{0}$-th content file is stored
in some of the BS caches, and the user scheduling probability is given
by $\Pr\left[S=1|\mathbf{K}=\mathbf{k},L_{0}=l_{0}\right]=1$.
\item $q_{l_{0}}=0$: The requested $l_{0}$-th content file does not exist
in the cache and has to be fetched from the backhaul. As more content
files need to be fetched from the backhaul (i.e., $\sum_{l\in\left\{ l\big|q_{l}=0\right\} }k_{l}$
increases), the user scheduling probability decreases.
\end{enumerate}
\item \textbf{Effect of $M$:} $M$ does not directly affect the user scheduling
probability. However, it indirectly affects the user scheduling probability
through the feasible region of the cache storage allocation vector
$\mathbf{q}$. As $M$ increases, a larger cache diversity can be
achieved by collectively caching more content files at all BSs (as
can be seen from the cache capacity constraint $\sum_{l=1}^{L}q_{l}\leq MB_{C}$).
Hence, the backhaul scheduling probability can be improved.
\end{enumerate}
\end{remrk}

Consider the case in Example \ref{example:backhaul-outage}, conditioned
on the typical user being served with the backhaul (i.e., $l_{0}\in\left\{ 5,6\right\} $
and $q_{l_{0}}=0$), the probability that the typical user being scheduled
for transmission is given by $\Pr\left[S=1|\mathbf{K}=\mathbf{k},L_{0}=l_{0}\right]=\frac{2}{5}$. 

\subsection{Conditional Physical Layer Successful Transmission Probability}

In the following, we analyze the physical layer successful transmission
probability conditioned on a given user request's realization and
loading of $B_{0}$. 
\begin{lemma}
[Conditional physical layer successful transmission probability]\label{lemma:conditional-phy-outage}
Conditioned on BS loading vector $\mathbf{k}$, the $l_{0}$-th content
file being requested by $u_{0}$, and $u_{0}$ being scheduled for
transmission (i.e., $S=1$), the physical layer successful transmission
probability is given by
\begin{equation}
\Pr\left[C\geq\tau\big|\mathbf{K}=\mathbf{k},L_{0}=l_{0},S=1\right]=\frac{\lambda_{l_{0}}^{b}/\lambda_{I}^{b}}{\lambda_{l_{0}}^{b}/\lambda_{I}^{b}+\beta\left(M,g_{0}\right)},\label{eq:phy-outage}
\end{equation}
where
\begin{equation}
g_{0}=\sum_{l\in\left\{ l\big|q_{l}\neq0\right\} }k_{l}+\min\left\{ \sum_{l\in\left\{ l\big|q_{l}=0\right\} }k_{l},B_{B}\right\} ,
\end{equation}
is the realization of $G_{0}$ (which represents the number of users
simultaneously served by $B_{0}$) conditioned on $\mathbf{K}=\mathbf{k}$,
\begin{equation}
\beta\left(M,g_{0}\right)=\frac{2}{\alpha}\left(2^{\frac{Mg_{0}\tau}{W}}-1\right)^{\frac{2}{\alpha}}B^{\prime}\left(\frac{2}{\alpha},1-\frac{2}{\alpha},2^{-\frac{Mg_{0}\tau}{W}}\right),\label{eq:beta}
\end{equation}
$B^{\prime}\left(x,y,z\right)\triangleq\int_{z}^{1}u^{x-1}\left(1-u\right)^{y-1}\mathrm{d}u$
is the complementary incomplete Beta function, 
\begin{equation}
\lambda_{l_{0}}^{b}=\begin{cases}
\lambda^{b}, & q_{l_{0}}=0,\\
\frac{q_{l_{0}}}{M}\lambda^{b} & q_{l_{0}}\neq0,
\end{cases}
\end{equation}
is the density of BSs that have access to the $l_{0}$-th content
file, and $\lambda_{I}^{b}=\lambda_{b}/M$ is the density of interfering
BS in the transmitting subband of $u_{0}$.
\end{lemma}

The proof can be found in Appendix \ref{subsec:Proof-of-Lemma-conditional-phy-outage}.
\begin{remrk}
Lemma \ref{lemma:conditional-phy-outage} shows the following effect
of the number of BS groups $M$ and the caching allocation strategy
\textbf{$\mathbf{q}$} on the conditional physical layer successful
transmission probability: 
\begin{enumerate}
\item \textbf{Effect of $M$ for given $\mathbf{q}$:} As $M$ increases,
the physical layer achievable rate decreases due to the lower spectral
efficiency caused by the smaller frequency reuse factor $1/M$. Hence,
the conditional physical layer successful transmission probability
decreases. As a result, there is a tradeoff between spectral efficiency
and user scheduling probability and we shall derive the optimal number
of subbands and BS groups $M$ to maximize the successful transmission
probability in Section \ref{sec:Optimization-of-Cache}. Note that
when there is only one subband and one BS group ($M=1$), each user
is always served by the geographically nearest BS, and the system
model reduces to the conventional cellular network with PPP distributed
BSs and users, as considered in \cite{yu2013downlink}. In this case,
$\lambda_{l_{0}}^{b}=\lambda_{I}^{b}=\lambda_{b}$, and the conditional
physical layer successful transmission probability degenerates to
$\Pr\left[C\geq\tau\big|\mathbf{K}=\mathbf{k},L_{0}=l_{0},S=1\right]=\frac{1}{1+\beta\left(1,g_{0}\right)}$,
which is consistent with the result in \cite{yu2013downlink}.
\item \textbf{Effect of $\mathbf{q}$ for given $M$:}
\begin{enumerate}
\item $q_{l_{0}}\neq0$: The file requested by $u_{0}$ is stored in the
cache of BSs, and the density of BSs that have access to the $l_{0}$-th
content file is given by $\lambda_{l_{0}}^{b}=\frac{q_{l_{0}}}{M}\lambda_{b}$.
As $q_{l_{0}}$ increases, the density of BSs that have access to
$l_{0}$-th content file also increases, and hence, the distance between
$u_{0}$ and $B_{0}$ decreases. Meanwhile, the interference of $u_{0}$
only comes from the BSs in the same group as $B_{0}$ with BS density
$\lambda_{I}^{b}=\frac{\lambda_{b}}{M}$, which is not affected by
$q_{l_{0}}$. As a result, when $q_{l_{0}}$ increases, the conditional
physical layer successful transmission probability increases. 
\item $q_{l_{0}}=0$: Since $u_{0}$ is associated with the geographically
nearest BS with BS density $\lambda_{l_{0}}^{b}=\lambda_{b}$, the
conditional physical layer successful transmission probability is
given by $\Pr\left[C\geq\tau\big|\mathbf{K}=\mathbf{k},L_{0}=l_{0},S=1\right]=\frac{M}{M+\beta\left(M,g_{0}\right)}$,
which is the same as the case when $q_{l_{0}}=M$.
\end{enumerate}
\end{enumerate}
\end{remrk}

\subsection{Average Successful Transmission Probability}

The average successful transmission probability is a function of the
BS loading $\mathbf{K}$, which is a random vector. To simplify the
analysis, we first compute the expectation of the BS loading vector
$\mathbf{\mathbf{K}}$ as follows.
\begin{lemma}
[Expectation of the BS loading vector $\mathbf{K}$]\label{theorem:Expectation-of-k}
The expectation of the BS loading vector $\mathbf{K}$ is given by
\begin{equation}
\tilde{k}_{l}\triangleq\mathbb{E}\left[K_{l}\right]=\rho_{l}+\frac{9\lambda_{u}\rho_{l}}{7\lambda_{b}}.\label{eq:expectation-k}
\end{equation}
\end{lemma}

The proof can be found in Appendix \ref{subsec:Proof-of-Theorem-expectation-k}. 

Now instead of considering the distribution of $\mathbf{K}$, we approximate
the BS loading $\mathbf{K}$ using its expectations in Lemma \ref{theorem:Expectation-of-k}.
The approximate successful transmission probability conditioned on
the $l_{0}$-th content file being requested by $u_{0}$ is given
by
\begin{align}
 & \Pr\left[C\geq\tau,\ S=1\big|L_{0}=l_{0}\right]\nonumber \\
= & \mathbb{E}_{\mathbf{K}}\Pr\left[S=1\big|\mathbf{K}=\mathbf{k},L_{0}=l_{0}\right]\nonumber \\
 & \times\Pr\left[C\geq\tau\big|\mathbf{K}=\mathbf{k},L_{0}=l_{0},S=1\right]\\
\approx & \Pr\left[S=1\big|\mathbf{K}=\mathbb{E}\left[\mathbf{K}\right],L_{0}=l_{0}\right]\nonumber \\
 & \times\Pr\left[C\geq\tau\big|\mathbf{K}=\mathbb{E}\left[\mathbf{K}\right],L_{0}=l_{0},S=1\right],\label{eq:approx-conditional-p}
\end{align}
where the approximation in (\ref{eq:approx-conditional-p}) is obtained
by replacing the probability density function of $\mathbf{K}$ with
a delta function $\delta\left(x-\mathbb{E}\left[\mathbf{K}\right]\right)$.

Using (\ref{eq:backhaul-outage}), (\ref{eq:phy-outage}), (\ref{eq:expectation-k}),
and (\ref{eq:approx-conditional-p}), the average successful transmission
probability can then be approximated as
\begin{align}
 & \tilde{p}\left(M,\mathbf{q}\right)\triangleq\sum_{l_{0}=1}^{L}\tilde{p}_{l_{0}}\rho_{l_{0}}\approx p\left(M,\mathbf{q}\right),\label{eq:approx-p}
\end{align}
where
\begin{equation}
\tilde{p}_{l_{0}}=\begin{cases}
\frac{q_{l_{0}}}{q_{l_{0}}+\beta\left(M,\tilde{g}_{0}\right)}, & q_{l_{0}}\neq0,\\
\frac{M}{M+\beta\left(M,\tilde{g}_{0}\right)}\min\left\{ \frac{B_{B}}{\sum_{l\in\left\{ l\big|q_{l}=0\right\} }\tilde{k}_{l}},1\right\} , & \mathrm{otherwise}.
\end{cases}\label{eq:approx-p-l0}
\end{equation}
and 
\begin{equation}
\tilde{g}_{0}=\sum_{l\in\left\{ l\big|q_{l}\neq0\right\} }\tilde{k}_{l}+\min\left\{ \sum_{l\in\left\{ l\big|q_{l}=0\right\} }\tilde{k}_{l},B_{B}\right\} .\label{eq:approx-loading}
\end{equation}
In Section \ref{sec:Simulation-Results}, simulations show that the
approximate gap between $\tilde{p}$ and the simulated successful
transmission probability is quite small under various scenarios. Therefore,
in the rest of the paper, the optimization of the frequency reuse
factor $1/M$ and cache storage capacity allocation vector $\mathbf{q}$
will be based on the approximate average successful transmission probability
in (\ref{eq:approx-p}). The accurate expression of the successful
transmission probability is also provided, but it is too complicated
to provide any useful insight. Interested readers should please refer
to Appendix \ref{subsec:Accurate-Expression-of-p} for details. In
the following section, we formulate and solve an optimization problem to find
the optimal $M$ and $\mathbf{q}$ that maximize the approximate average
successful transmission probability.

\section{Optimization of Cache Storage Allocation and Frequency Reuse\label{sec:Optimization-of-Cache}}

\subsection{Problem Formulation}

The problem of finding the optimal frequency reuse factor and cache
storage capacity allocation vector that maximize the approximate average
successful transmission probability is formulated as:
\begin{align}
\mathcal{P}:\max_{M\in\mathbb{N}^{+},q_{l}\in\left\{ 0,1,\dots,M\right\} ,\forall l}\  & \tilde{p}\left(M,\mathbf{q}\right)\\
\mathrm{s.t.}\qquad\quad\; & q_{l}\geq q_{l+1},l=1,\dots,L-1,\label{eq:property-q-order}\\
 & \sum_{l=1}^{L}q_{l}\leq MB_{C}.
\end{align}
Denote $M^{\star}$ and $\mathbf{q}^{\star}$ as the optimal solution
of $\mathcal{P}$. Constraint (\ref{eq:property-q-order}) is used
to simplify the optimization algorithm design. Simulation results
show that our proposed scheme with constraint (\ref{eq:property-q-order})
achieves a reasonably large average successful transmission probability
gain over existing caching schemes in typical scenarios.

Problem $\mathcal{P}$ is an integer optimization problem, and the
objective function is neither convex nor concave. Even if we fix $M$
and relax the integer constraint on $\mathbf{q}$ to allow it to be
a real vector, the relaxed problem is still very difficult to solve
due to the indicator function w.r.t. $q_{l}$ in (\ref{eq:approx-p-l0})
and (\ref{eq:approx-loading}), and the complicated function $\beta\left(M,\tilde{g}_{0}\right)$
w.r.t. $\tilde{g}_{0}$ (recall that $\tilde{g}_{0}$ also depends
on $\mathbf{q}$) in (\ref{eq:beta}). As a result, it is highly non-trivial
to even design a low-complexity algorithm for Problem $\mathcal{P}$
by solving the above relaxed problem. 

\subsection{Problem Transformation and Optimization}

For a fixed $M$, the primary difficulty in solving Problem $\mathcal{P}$
is how to deal with the user scheduling probability (\ref{eq:backhaul-outage})
and expectation of the BS loading (\ref{eq:expectation-k}), in which
$\mathbf{q}$ appears in the indication function in the subscript
of summation. To address this challenge, we introduce an auxiliary
variable $L^{\prime}$, which is the number of content files that
can be found in BS caches. Under (\ref{eq:property-q-order}), we
have
\begin{align}
q_{l}\geq1 & ,\ \forall l\leq L^{\prime},\\
q_{l}=0 & ,\ \forall l>L^{\prime}.
\end{align}
Note that for a given $L^{\prime}$, the set of content files that
need to be fetched from backhaul (whose indexes are given by $\left\{ L^{\prime}+1,\dots,L\right\} $)
is fixed. As a result, the backhaul success probability given by (\ref{eq:backhaul-outage})
is fixed. If we further assume $M$ is given, then for the content
files indexed by $\left\{ L^{\prime}+1,\dots,L\right\} $, the successful
transmission probability is also fixed. In this case, to find the
optimal $\mathbf{q}$ that maximizes the average successful transmission
probability, we only need to minimize the average physical layer outage
probability over the content files that are stored in BS caches. Specifically,
after relaxing the integer constraint on $\mathbf{q}$, $\mathcal{P}$
can be decomposed into a set of sub-problems that minimize the average
physical layer outage probability for a given $M$ and $L^{\prime}$,
which is given by
\begin{align}
\tilde{\mathcal{P}}\left(M,L^{\prime}\right):\min_{q_{l}} & \sum_{l=1}^{L^{\prime}}\frac{\rho_{l}\beta\left(M,\tilde{g}_{0}\right)}{q_{l}+\beta\left(M,\tilde{g}_{0}\right)}\label{eq:sub-problems}\\
\mathrm{s.t.}\  & \sum_{l=1}^{L^{\prime}}q_{l}=MB_{C},\\
 & q_{l}\geq q_{l+1},\ \forall l\in\left\{ 1,\dots,L^{\prime}-1\right\} ,\label{eq:ql-order}\\
 & 1\leq q_{l}\leq M,\ \forall l\in\left\{ 1,\dots,L^{\prime}\right\} 
\end{align}
for $M\in\mathbb{N}^{+}$ and $L^{\prime}\in\left[B_{C},\min\left\{ MB_{C},L\right\} \right]$.
Denote $\tilde{\mathbf{q}}^{\star}\left(M,L^{\prime}\right)$ as the
optimal solution of the sub-problem $\tilde{\mathcal{P}}\left(M,L^{\prime}\right)$.
Note that for the given $M$ and $L^{\prime}$, both $\tilde{K}_{0}$
and $\beta\left(M,\tilde{g}_{0}\right)$ are fixed. It can be easily
seen that $\tilde{\mathcal{P}}\left(M,L^{\prime}\right)$ is a convex
minimization problem, and $\tilde{\mathbf{q}}^{\star}\left(M,L^{\prime}\right)$
can be obtained using Karush\textendash Kuhn\textendash Tucker (KKT)
conditions \cite{boyd2004convex}, as in the following theorem.
\begin{thm}
[Optimal solution of $\tilde{\mathcal{P}}\left(M,L^{\prime}\right)$]\label{thm:Optimal-solution-sub}
The optimal solution $\tilde{\mathbf{q}}^{\star}\left(M,L^{\prime}\right)$
of problem $\tilde{\mathcal{P}}\left(M,L^{\prime}\right)$ is given
by
\begin{align}
 & \tilde{q}_{l}^{\star}\left(M,L^{\prime}\right)=\min\bigg\{ M,\nonumber \\
 & \max\left\{ 1,\sqrt{\rho_{l}/\lambda^{\star}}-\beta\left(M,\tilde{g}_{0}\right)\right\} \bigg\},\forall l\in\left\{ 1,\dots,L^{\prime}\right\} ,
\end{align}
where $\lambda^{\star}$ satisfies
\begin{equation}
\sum_{l=1}^{L^{\prime}}\min\left\{ M,\max\left\{ 1,\sqrt{\rho_{l}/\lambda^{\star}}-\beta\left(M,\tilde{g}_{0}\right)\right\} \right\} =MB_{C}.
\end{equation}
\end{thm}

The proof can be found in Appendix \ref{subsec:Proof-of-Theorem-optimal-solution}. 

The content popularity distribution $\rho$ and physical layer parameter
represented by $\beta\left(M,\tilde{g}_{0}\right)$ jointly affect
$\tilde{\mathbf{q}}^{\star}\left(M,L^{\prime}\right)$. Note that
content file with higher popularity is allocated more cache storage
resources. For a heavy-tailed popularity distribution, the differences
between $\rho_{l}$'s are small, and the cache capacity is allocated
to more content files instead of concentrating on a few most popular
files.

To calculate a solution of $\mathcal{P}$, we first enumerate $L^{\prime}$
and $M$ to find the best solution $\left(\tilde{M}^{\star},\tilde{\mathbf{q}}^{\star}\left(\tilde{M}^{\star},L^{\prime\star}\right)\right)$
that maximizes the objective function $\tilde{p}\left(M,\mathbf{q}\right)$
of $\mathcal{P}$. Then $\tilde{\mathbf{q}}^{\star}\left(\tilde{M}^{\star},L^{\prime\star}\right)$
is rounded down such that it becomes a feasible integer solution of
$\mathcal{P}$. Finally, the residue cache storage capacity induced
by the rounding down operation is allocated to the content files that
minimizes the average physical layer outage probability in a greedy
manner. The detailed algorithm is given in Algorithm \ref{alg:Subbands-decision-and-cache-allocation}.
Note that the optimal objective value $\tilde{p}\left(\tilde{M}^{\star},\tilde{\mathbf{q}}^{\star}\left(M^{\star},L^{\prime\star}\right)\right)$
for the relaxed problem provides an upper bound of the optimal objective
value of the original problem $\mathcal{P}$. In Section \ref{sec:Simulation-Results},
we show that this upper bound is quite close to the objective
value achieved by the integer solution $\left(\tilde{M}^{\star},\hat{\mathbf{q}}^{\star}\left(M^{\star},L^{\prime\star}\right)\right)$
obtained using Algorithm \ref{alg:Subbands-decision-and-cache-allocation},
which shows that the proposed low-complexity algorithm is close-to-optimal
for the original integer optimization problem. 

In practice, we can set a limit $M_{max}$ for the maximum number
of subbands searched by Algorithm \ref{alg:Subbands-decision-and-cache-allocation},
and $M_{max}$ can be used to control the tradeoff between performance
and complexity. Note that the optimal number of subbands $\tilde{M}^{\star}$
is usually small. Otherwise the bandwidth would become insufficient
to support the transmission rate $\tau$. $L^{\prime}$ is upper bounded
by $MB_{C}$, which is usually much smaller than the total number
of content files. Algorithm \ref{alg:Subbands-decision-and-cache-allocation}
needs to solve $M_{\max}^{2}B_{C}$ sub-problems (\ref{eq:sub-problems}),
and each sub-problem can be efficiently solved using bisection. Note
that the cache capacity $B_{C}$ is usually much smaller than the
total number of content files $L$. As a result, the computational
complexity induced by enumeration is low. In Section \ref{sec:Simulation-Results},
we shall show that Algorithm \ref{alg:Subbands-decision-and-cache-allocation}
is quite efficient and achieves a large average successful transmission
probability gain over conventional single-band caching schemes. In
Section \ref{sec:Simulation-Results}, we shall also provide numerical
insight on $\tilde{M}^{\star}$ and $L^{\prime\star}$, i.e., the
optimal number of subbands and BS groups, and how many content files
to cache.

\begin{algorithm}[t]
\caption{\label{alg:Subbands-decision-and-cache-allocation}Subbands decision
and cache storage capacity allocation}

\begin{algorithmic}[1]

\FOR{$M=1,\dots,M_{\max}$}

\FOR{$L^{\prime}=B_{C},\dots,\min\left\{ MB_{C},L\right\} $}

\STATE Calculate optimal solution $\tilde{\mathbf{q}}^{\star}\left(M,L^{\prime}\right)$
of problem $\tilde{\mathcal{P}}\left(M,L^{\prime}\right)$ using Theorem
\ref{thm:Optimal-solution-sub}

\ENDFOR

\ENDFOR

\STATE $\left(\tilde{M}^{\star},L^{\prime\star}\right)=\arg\max\tilde{p}\left(M,\mathbf{q}\right)$
and $\tilde{\mathbf{q}}^{\star}=\tilde{\mathbf{q}}^{\star}\left(\tilde{M}^{\star},L^{\prime\star}\right)$

\STATE $\hat{\mathbf{q}}^{\star}=\left\lfloor \tilde{\mathbf{q}}^{\star}\right\rfloor $,
where $\left\lfloor \cdot\right\rfloor $ is the rounding down function

\WHILE{$\sum_{l=1}^{L}\hat{q}_{l}^{\star}<\tilde{M}^{\star}B_{C}$}

\STATE $l^{\prime}=\arg\min_{l\in\left\{ 1,\dots,L^{\prime\star}\right\} }\left(\frac{\rho_{l}}{\hat{q}_{l}^{\star}+1+\beta\left(\tilde{M}^{\star},\tilde{g}_{0}\right)}-\frac{\rho_{l}}{\hat{q}_{l}^{\star}+\beta\left(\tilde{M}^{\star},\tilde{g}_{0}\right)}\right)$

\STATE $\hat{q}_{l^{\prime}}^{\star}=\hat{q}_{l^{\prime}}^{\star}+1$

\ENDWHILE

\end{algorithmic}
\end{algorithm}

\section{Simulation Results\label{sec:Simulation-Results}}

In this section, we show that the simulation results are consistent
with the theoretical results. We also demonstrate the performance
gain of our scheme over the following baselines: 
\begin{itemize}
\item \textbf{Baseline 1:} A standard policy which caches the most popular
content (MPC) at each BS \cite{bacstug2015cache}. The frequency reuse
factor is one, and orthogonal frequency-division multiple access (OFDMA)
is applied at each BS to serve multiple users. 
\item \textbf{Baseline 2:} Geographic caching problem (GCP) proposed in
\cite{blaszczyszyn2015optimal}, which exploits spatial caching diversity
by random caching. The frequency reuse factor is one, and OFDMA is
applied at each BS to serve multiple users. The corresponding cache
placement is optimized to maximize the average successful transmission
probability (the performance metric considered in this paper), using
an algorithm similar to Algorithm \ref{alg:Subbands-decision-and-cache-allocation},
which combines enumeration and convex optimization.
\item \textbf{Baseline 3:} Separated design of MPC and frequency reuse.
The frequency reuse factor is optimized to maximize the average successful
transmission probability.
\item \textbf{Baseline 4:} Separated design of GCP and frequency reuse.
The cache placement and frequency reuse factor are separately optimized
to maximize the average successful transmission probability.
\end{itemize}
For the proposed scheme, the number of subbands $\tilde{M}^{\star}$
and cache storage allocation vector $\hat{\mathbf{q}}^{\star}$ is
obtained using Algorithm \ref{alg:Subbands-decision-and-cache-allocation},
where the maximum number of subbands is given by $M_{\max}=5$. The
system parameters are set as follows:
\begin{itemize}
\item \textbf{Geometric parameters:} BS density $\lambda^{b}=3\times10^{-5}$,
user density $\lambda^{u}=3\times10^{-4}$.
\item \textbf{Channel parameters:} Path loss exponent $\alpha=4$, total
bandwidth $W=20$MHz, target rate $\tau=0.1$ Mbps, length of each
time slot $\nu=1$ ms.
\item \textbf{Content parameters:} Content library size $L=1000$, content
popularity follows Zipf distribution with exponent $\gamma$ \cite{olivier2013performance,yamakami2006zipf}.
\end{itemize}
\begin{figure}[t]
\begin{centering}
\includegraphics[width=1\columnwidth]{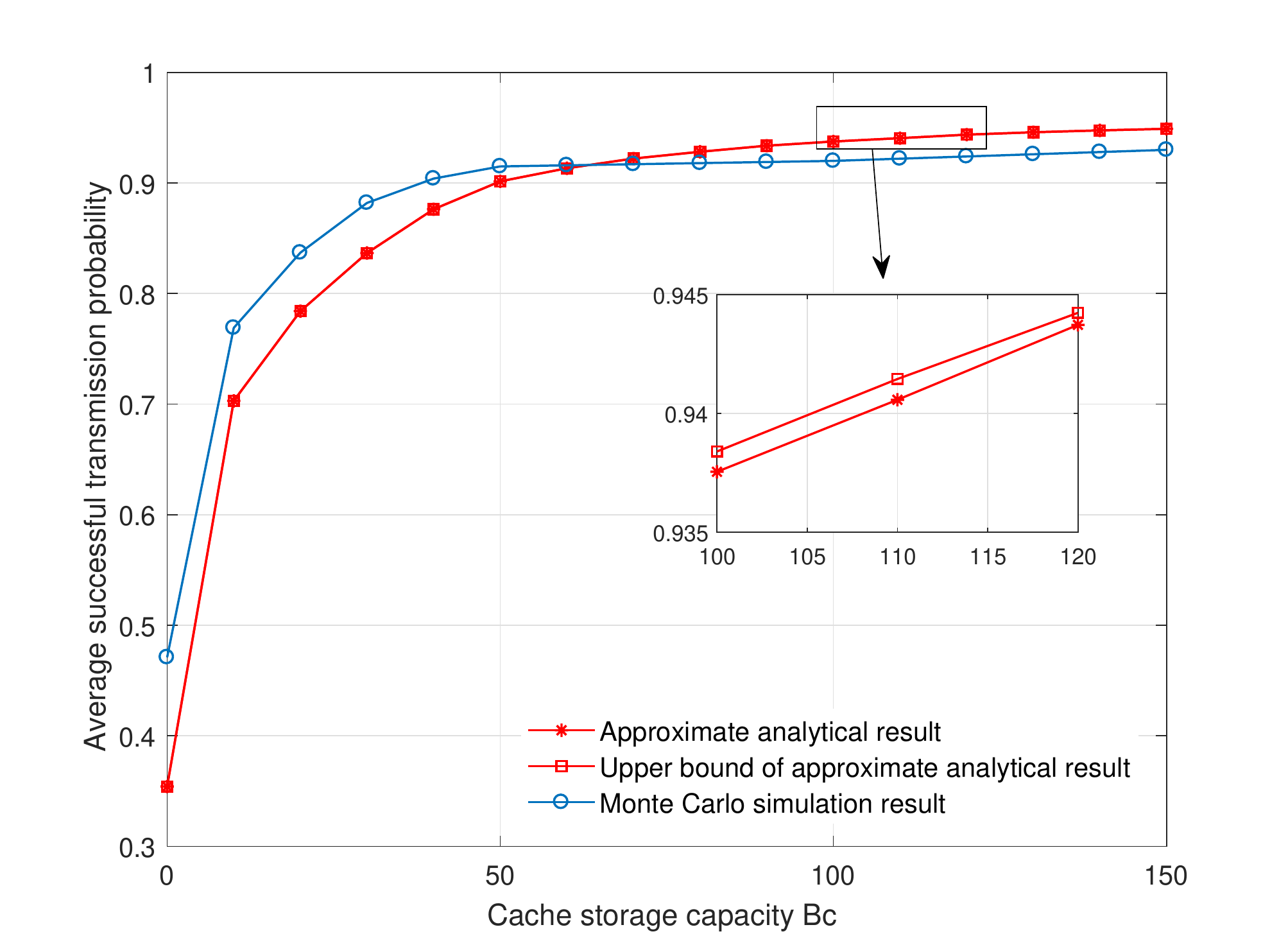}
\par\end{centering}
\caption{\label{fig:Illustration-of-the-approximation}Illustration of the
approximate gap between $\tilde{p}$ and $p$ when $\gamma=0.8$ and
$B_{B}=5$.}
\end{figure}

In Fig. \ref{fig:Illustration-of-the-approximation}, we plot the
average successful transmission probability versus the cache storage
capacity when $\gamma=0.8$ and $B_{B}=5$. It can be observed that
the simulation results largely match the theory; the relative approximation
error is large only when the average successful transmission probability
is very low, which is not a desirable operating regime for practical
systems. In a practical regime when the success probability is high,
the approximate error is small. In this case, the theoretical approximation
(\ref{eq:approx-p}) can capture the first-order behavior of the proposed
scheme and can be used to optimize the cache design. The results also
show that the objective value achieved using Algorithm \ref{alg:Subbands-decision-and-cache-allocation}
is quite close to the upper bound given by $\tilde{p}\left(\tilde{M}^{\star},\tilde{\mathbf{q}}^{\star}\left(M^{\star},L^{\prime\star}\right)\right)$
(it is an upper bound since the integer constraint on $\mathbf{q}$
is relaxed), which indicates that the proposed low-complexity algorithm
is close to optimum for the original integer optimization problem.

\begin{figure}[t]
\begin{centering}
\includegraphics[width=1\columnwidth]{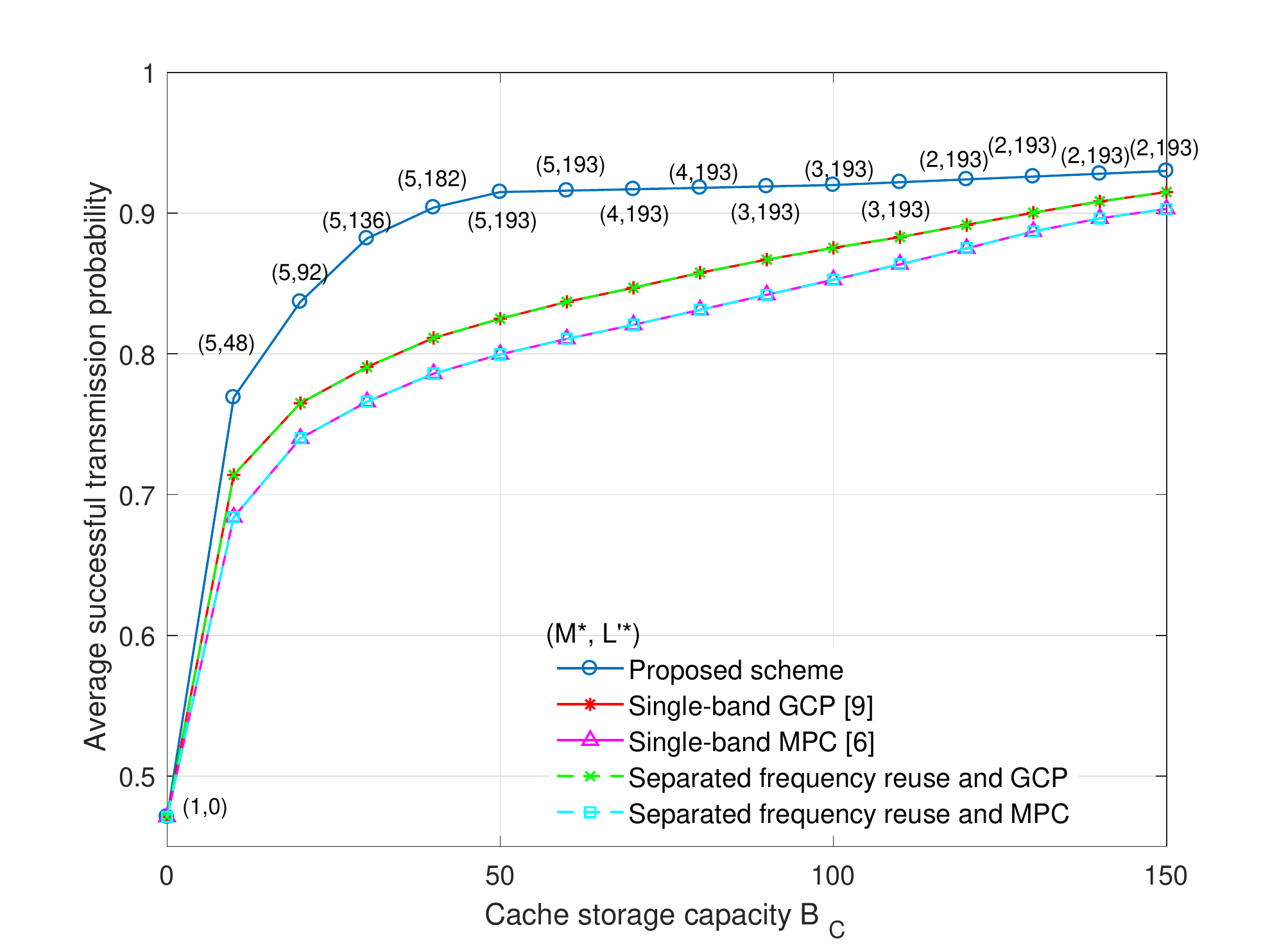}
\par\end{centering}
\caption{\label{fig:p-bc}Average successful transmission probability versus
cache capacity when $\gamma=0.8$ and $B_{B}=5$.}
\end{figure}

\begin{figure}[t]
\begin{centering}
\includegraphics[width=1\columnwidth]{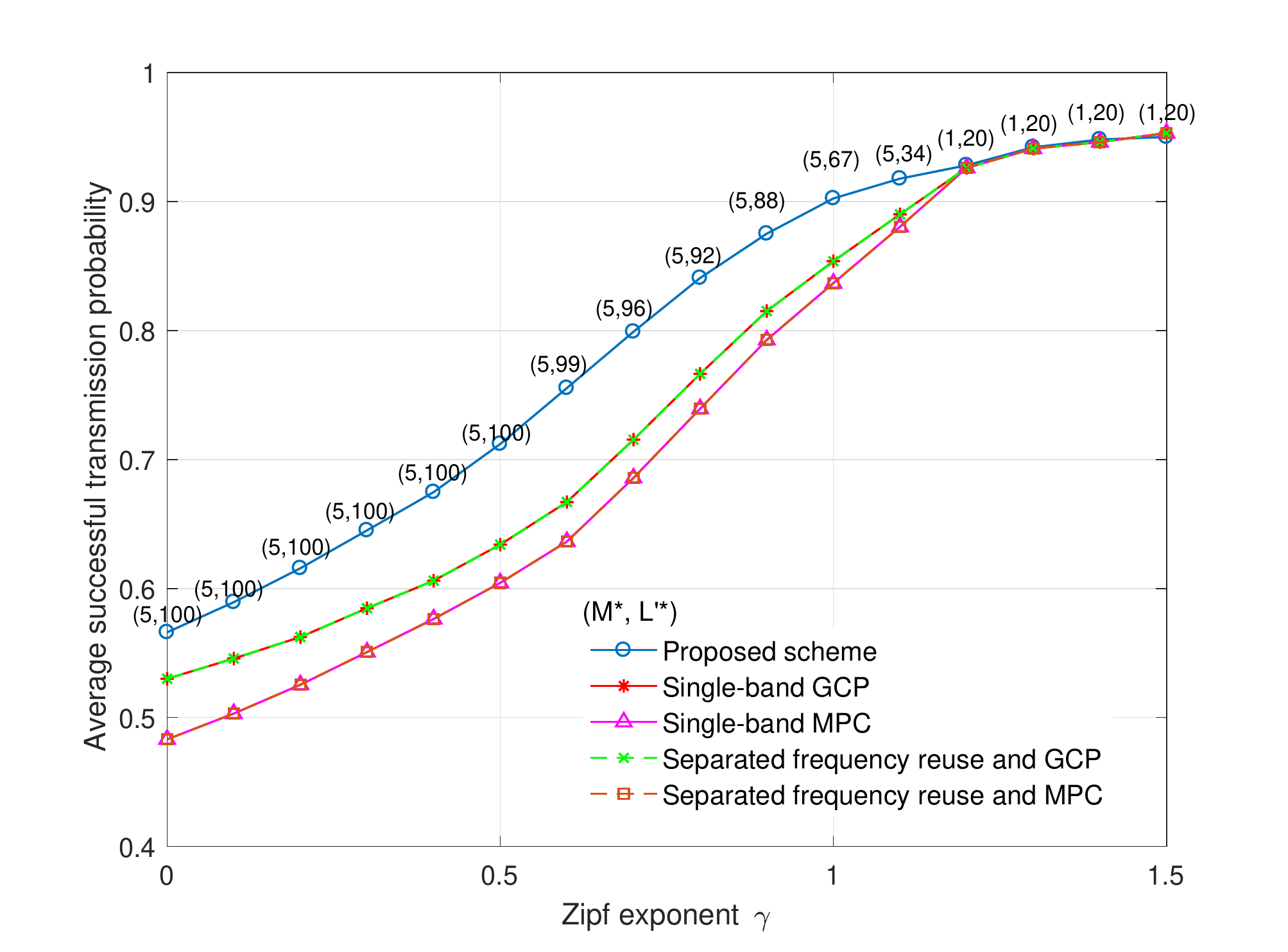}
\par\end{centering}
\caption{\label{fig:p-gamma}Average successful transmission probability versus
Zipf distribution exponent when $B_{C}=20$ and $B_{B}=5$.}
\end{figure}

In Fig. \ref{fig:p-bc} \textendash{} Fig. \ref{fig:p-gamma}, we
compare the performance between the proposed scheme and the baselines.
It is observed that for separated design and optimization of frequency
reuse with either GCP or MPC, the optimal frequency reuse strategy
is usually to let all BSs use the entire bandwidth (i.e., optimal
frequency reuse factor is one). This is because a separated design
of frequency reuse and GCP cannot completely address the strong interference
issue in random caching. Meanwhile, for separated design and optimization
of frequency reuse and MPC, each scheduled user is always served by
the geographically nearest BS, and thus the inter-cell interference
is weaker compared to the case with random caching. In both cases,
a frequency reuse factor less than one would lead to lower physical
layer successful transmission probability due to less bandwidth being
allocated to each BS.

\begin{itemize}
\item \textbf{Impact of cache capacity $B_{C}$ (Fig. \ref{fig:p-bc}):}
\begin{itemize}
\item \textbf{On the optimal number of subbands $\tilde{M}^{\star}$:} When
$B_{C}$ is small, the optimal number of subbands $\tilde{M}^{\star}$
is large, so that the proposed scheme can achieve a lower backhaul
outage probability by exploiting spatial caching diversity. As $B_{C}$
increases, due to sufficient cache storage capacity, the importance
of spatial caching diversity decreases. As a result, $\tilde{M}^{\star}$
decreases, so that the distance between the user and BS decreases
and the physical layer achieves larger spectral efficiency. 
\item \textbf{On the optimal caching strategy:} When $B_{C}$ is small,
the optimal cache storage allocation of the proposed scheme is not
to use the cache to store only the most popular content files in every
BS (i.e., $L^{\prime\star}\neq B_{C}$). As $B_{C}$ increases, $L^{\prime\star}$
increases since more content files can be stored in the cache, which
leads to a lower backhaul outage probability. 
\end{itemize}
\end{itemize}
It can be seen that single-band MPC \cite{bacstug2015cache} and single-band
GCP \cite{blaszczyszyn2015optimal} are not optimal when the cache
capacity $B_{C}$ is limited.

\begin{figure}[t]
\begin{centering}
\includegraphics[width=1\columnwidth]{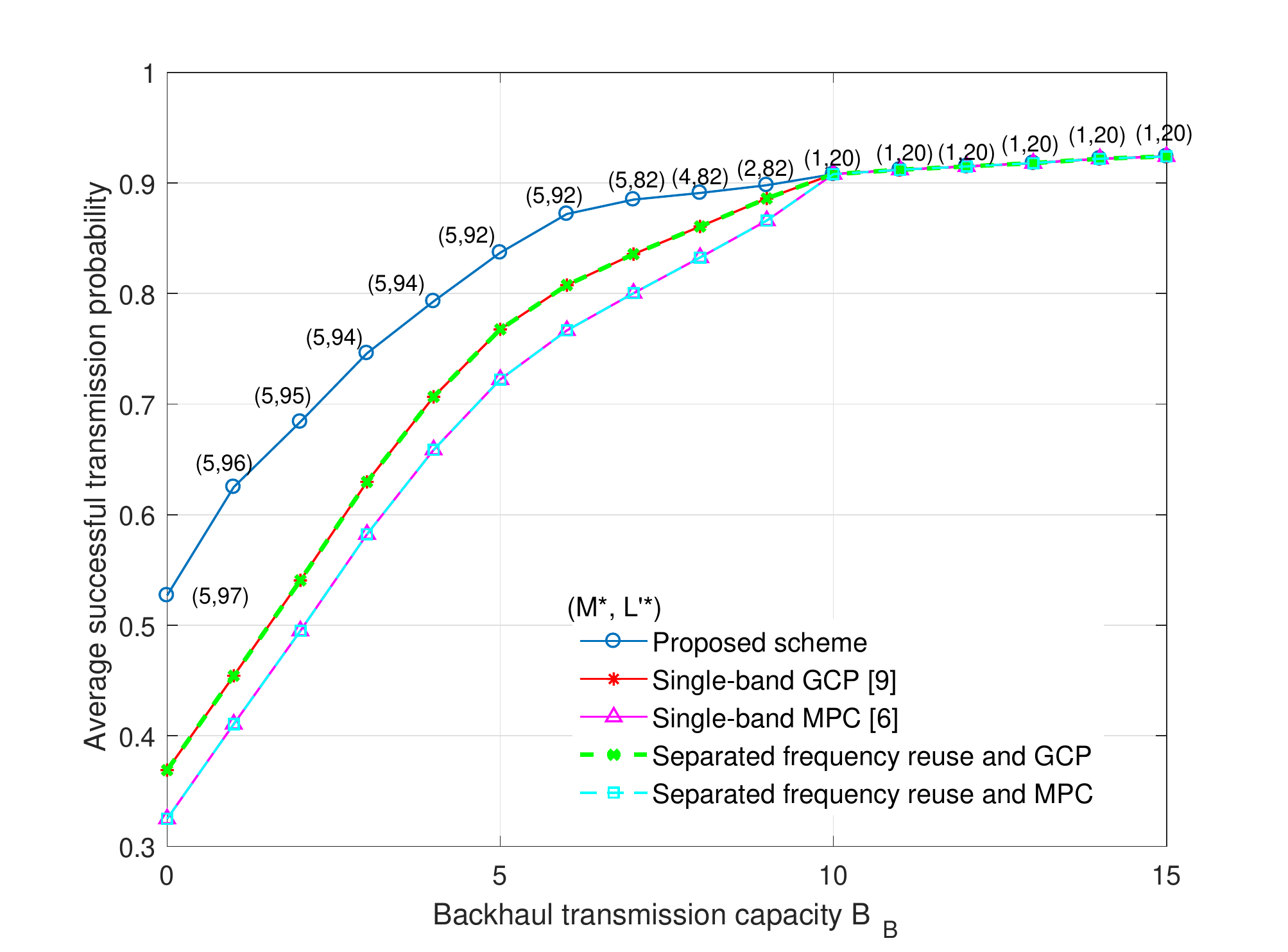}
\par\end{centering}
\caption{\label{fig:p-bb}Average successful transmission probability versus
backhaul transmission capacity when $\gamma=0.8$ and $B_{C}=20$.}
\end{figure}

\begin{itemize}
\item \textbf{Impact of backhaul transmission capacity $B_{B}$ (Fig. \ref{fig:p-bb}): }
\begin{itemize}
\item \textbf{On the optimal number of subbands $\tilde{M}^{\star}$:} When
the backhaul capacity $B_{B}$ is small, the backhaul can handle fewer
user requests. In this case, it is better to increase $M$ to improve
the spatial caching diversity gain and cache hit probability. When
$B_{B}$ is large, the backhaul can handle more user requests and
the importance of spatial caching diversity decreases. In this case,
it is better to decrease $M$ to achieve larger physical layer spectral
efficiency. Therefore, the optimal number of subbands $\tilde{M}^{\star}$
decreases with the cache capacity $B_{B}$.
\item \textbf{On the optimal caching strategy:} When cache capacity $B_{B}$
is small, it is important to exploit the spatial caching diversity
to improve the cache hit probability. In this case, the optimal cache
storage allocation is not to use the cache to store only the most
popular content files in every BS, but to use some cache capacity
to store some less popular content files as well. On the other hand,
as $B_{B}$ increases, more content files can be fetched from the
backhaul without causing backhaul outage. In this case, it is not
necessary to store many less popular content files; i.e., $L^{\prime\star}$
will decrease.
\end{itemize}
\end{itemize}
It can be seen that single-band MPC \cite{bacstug2015cache} and single-band
GCP \cite{blaszczyszyn2015optimal} are not optimal when the backhaul
transmission capacity is limited.

\begin{itemize}
\item \textbf{Impact of }Zipf exponents $\gamma$\textbf{ (Fig. }\ref{fig:p-gamma}\textbf{):}
\begin{itemize}
\item \textbf{On the optimal number of subbands $\tilde{M}^{\star}$:} When
$\gamma$ is small (i.e., the popularity distribution is flat), it
is important to exploit the spatial caching diversity to improve the
cache hit probability. In this case, it is better to increase $M$
to improve the spatial caching diversity gain and cache hit probability.
On the other hand, as $\gamma$ increases, the user requests concentrate
on a few content files, and hence the benefit of spatial caching diversity
decreases. Therefore, the optimal number of subbands $\tilde{M}^{\star}$
decreases.
\item \textbf{On the optimal caching strategy:} When $\gamma$ is small,
it is important to exploit the spatial caching diversity to improve
the cache hit probability. In this case, the optimal cache storage
allocation is not to use the cache to store only the most popular
content files in every BS. As $\gamma$ increases, the user requests
concentrate on a few content files, hence the benefit of spatial caching
diversity decreases. In this case, it is not desirable to store too
many less popular content files, i.e., $L^{\prime\star}$ will decrease.
\end{itemize}
\end{itemize}
Based on the simulation results, our proposed scheme outperforms single-band
MPC \cite{bacstug2015cache} and single-band GCP \cite{blaszczyszyn2015optimal},
especially for the values of $\gamma$ from 0 to 1, with is typical
for general applications \cite{olivier2013performance}. 

\begin{figure}[t]
\begin{centering}
\includegraphics[width=1\columnwidth]{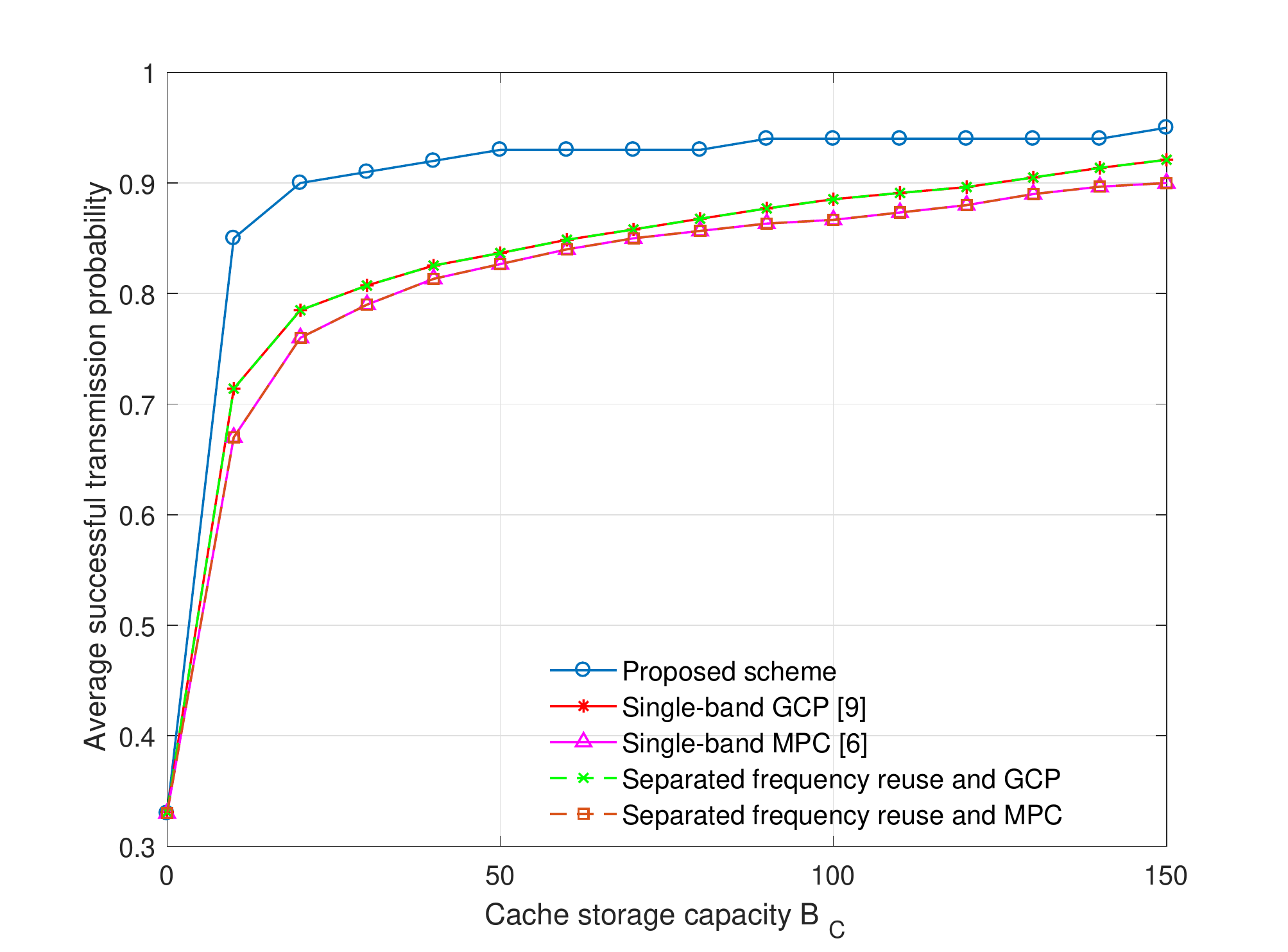}
\par\end{centering}
\caption{\label{fig:fixed-bc-p}Successful transmission probability versus
cache capacity under a given realization of PPP when $\gamma=0.8$
and $B_{B}=5$.}
\end{figure}

Additionally, in Fig. \ref{fig:fixed-bc-p}, we plot the successful
transmission probability under a given realization of PPP, which shows
that the proposed design also works well in this case. 

\section{Conclusion\label{sec:Conclusion}}

In this paper, we propose a joint frequency reuse and caching scheme
to achieve both the spatial cache diversity and interference mitigation
in small-cell backhaul-limited wireless networks. We first derive
a closed-form expression of the approximate successful transmission
probability using the tools of stochastic geometry, and analyze the
impact of key operating parameters. We then propose a low-complexity
algorithm which combines enumeration and convex optimization to optimize
the frequency reuse factor and the cache storage allocation vector.
Finally, by simulations, we show that by exploiting the spatial cache
diversity and interference mitigation benefits provided by the joint
optimization of frequency reuse and caching, the proposed scheme achieves
a large performance gain over the typical single-band MPC scheme \cite{bacstug2015cache}
and random caching scheme \cite{blaszczyszyn2015optimal}, especially
when the cache capacity and backhaul capacity at each BS are limited.

\appendix

\subsection{Proof of Lemma \ref{lemma:conditional-phy-outage} \label{subsec:Proof-of-Lemma-conditional-phy-outage}}

First, we calculate the  physical layer successful transmission probability
conditioned on file $l_{0}$ requested by $u_{0}$, BS loading $\mathbf{K}=\mathbf{k}$,
$u_{0}$ being scheduled for transmission (i.e., $S=1$) and distance
$D_{0,0}=d$, which is given by
\begin{align}
 & \Pr\left[C\geq\tau\big|\mathbf{K}=\mathbf{k},L_{0}=l_{0},S=1,D_{0,0}=d\right]\nonumber \\
 & \overset{\left(a\right)}{=}\mathbb{E}_{\Phi_{m_{0}}^{b}}\left[\exp\left(-\left(2^{\frac{Mg_{0}\tau}{W}}-1\right)d^{\alpha}\sum_{n\in\Phi_{m_{0}}^{b}\backslash B_{0}}D_{n,0}^{-\alpha}\left|h_{n,0}\right|^{2}\right)\right]\nonumber \\
 & =\mathbb{E}_{\Phi_{m_{0}}^{b}}\left[\prod_{n\in\Phi_{m_{0}}^{b}\backslash B_{0}}\exp\left(-\left(2^{\frac{Mg_{0}\tau}{W}}-1\right)d^{\alpha}D_{n,0}^{-\alpha}\left|h_{n,0}\right|^{2}\right)\right]\nonumber \\
 & \overset{\left(b\right)}{=}\exp\left(-2\pi\lambda_{I}^{b}\int_{d}^{\infty}\left(1-\frac{1}{1+\left(2^{\frac{Mg_{0}\tau}{W}}-1\right)d^{\alpha}r^{-\alpha}}\right)r\mathrm{d}r\right)\nonumber \\
 & \overset{\left(c\right)}{=}\exp\left(-\frac{2\pi}{\alpha}\lambda_{I}^{b}\left(2^{\frac{Mg_{0}\tau}{W}}-1\right)^{\frac{2}{\alpha}}B^{\prime}\left(\frac{2}{\alpha},1-\frac{2}{\alpha},2^{-\frac{Mg_{0}\tau}{W}}\right)d^{2}\right)\nonumber \\
 & =1\exp\left(-\pi\lambda_{I}^{b}\beta\left(M,g_{0}\right)d^{2}\right),\label{eq:conditioned-phy-outage-d}
\end{align}
where (a) is due to Rayleigh fading channel $\left|h_{0,0}\right|^{2}\overset{d}{\sim}\mathrm{Exp}\left(1\right)$,
(b) is obtained using the probability generating function of PPP \cite[Page 235]{haenggi2009interference},
and (c) is obtained by replacing $\left(2^{\frac{Mg_{0}\tau}{W}}-1\right)^{-\frac{1}{\alpha}}d^{-1}r$
with $t$, and then replacing $\frac{1}{1+t^{-\alpha}}$ with $w$. 

Then, we calculate $\Pr\left[C<\tau\big|\mathbf{K}=\mathbf{k},L_{0}=l_{0},S=1\right]$
by removing the condition of $D_{0,0}=d$. The probability density
function of $D_{0,0}$ is given by 
\begin{equation}
f_{D_{0,0},l_{0}}\left(d\right)=2\pi\lambda_{l_{0}}^{b}d\exp\left(-\pi\lambda_{l_{0}}^{b}d^{2}\right),\label{eq:distribution-d}
\end{equation}
as the BSs storing the $l_{0}$-th content file form a homogeneous
PPP with density $\lambda_{l_{0}}^{b}$. By (\ref{eq:conditioned-phy-outage-d})
and (\ref{eq:distribution-d}), we have
\begin{align}
 & \Pr\left[C\geq\tau\big|l_{0},K_{0},S=1\right]\nonumber \\
= & \int_{0}^{\infty}\Pr\left[C<\tau\big|\mathbf{K}=\mathbf{k},L_{0}=l_{0},S=1,D_{0,0}=d\right]\nonumber \\
 & \times f_{D_{0,0},l_{0}}\left(d\right)\mathrm{d}d\\
= & 2\pi\lambda_{l_{0}}^{b}\int_{0}^{\infty}d\exp\left(-\pi\left(\lambda_{l_{0}}^{b}+\lambda_{I}^{b}\beta\left(M,g_{0}\right)\right)d^{2}\right)\mathrm{d}d\\
\overset{\left(a\right)}{=} & \frac{\frac{\lambda_{l_{0}}^{b}}{\lambda_{I}^{b}}}{\frac{\lambda_{l_{0}}^{b}}{\lambda_{I}^{b}}+\beta\left(M,g_{0}\right)},
\end{align}
where (a) is obtained using $\int_{0}^{\infty}d\exp\left(-cd^{2}\right)\mathrm{d}d=\frac{1}{2c}$
($c$ is a constant).

\subsection{Proof of Lemma \ref{theorem:Expectation-of-k} \label{subsec:Proof-of-Theorem-expectation-k}}

In the following, we first derive the probability mass function of
$\mathbf{k}$. Note that due to the content-centric user scheduling
scheme, each file $X_{l}\in\left\{ X_{1},X_{2},\dots,X_{L}\right\} $
corresponds to a Voronoi tessellation, which is determined by the
locations of all BSs which have access to file $X_{l}$. To calculate
the probability mass function of $\mathbf{k}$, we need the probability
density function of the size of the Voronoi cell which $B_{0}$ belongs
to. Based on a widely used approximated form of this probability density
function given in \cite{yu2013downlink}, the probability mass function
of $\mathbf{K}$ is given in the following lemma. 
\begin{lemma}
[Probability mass function of $\mathbf{K}$]\label{lemma:pmf-k}
The probability mass function of $\mathbf{K}$ conditioned on the
$l_{0}$-th content file being requested by $u_{0}$ is given by 
\begin{align}
 & \Pr\left[\mathbf{K}=\mathbf{k}\big|L_{0}=l_{0}\right]\nonumber \\
 & =\frac{1}{q_{l_{0}}}\sum_{m_{0}\in\mathcal{M}_{l_{0}}}\Pr\left[\mathbf{K}=\mathbf{k}\big|L_{0}=l_{0},M_{0}=m_{0}\right]\label{eq:pmf-k-1}
\end{align}
for $q_{l_{0}}\neq0$, and 
\begin{align}
 & \Pr\left[\mathbf{K}=\mathbf{k}\big|L_{0}=l_{0}\right]\nonumber \\
 & =\frac{1}{M}\sum_{m_{0}\in\left\{ 0,\dots,M-1\right\} }\Pr\left[\mathbf{K}=\mathbf{k}\big|L_{0}=l_{0},M_{0}=m_{0}\right]\label{eq:pmf-k-2}
\end{align}
for $q_{l_{0}}=0$, where $\mathcal{M}_{l_{0}}$ is the set of indexes
of BS groups that stores the $l_{0}$-th content file, $\Pr\left[\mathbf{K}=\mathbf{k}\big|L_{0}=l_{0},M_{0}=m_{0}\right]$
is the probability mass function of $\mathbf{K}$ conditioned on the
$l_{0}$-th content file being requested by $u_{0}$ and $B_{0}\in\Phi_{m_{0}}^{b}$,
and $\Pr\left[K_{l}=k_{l}\big|L_{0}=l_{0},M_{0}=m_{0}\right]$ is
given by
\begin{align}
 & \Pr\left[K_{l}=k_{l}\big|L_{0}=l_{0},M_{0}=m_{0}\right]\nonumber \\
 & =\begin{cases}
\Psi\left(\lambda^{u}\rho_{l},\frac{q_{l}\lambda^{b}}{M}\right), & l=l_{0},q_{l_{0}}\neq0,\\
\Psi\left(\lambda^{u}\rho_{l},\lambda^{b}\right), & l=l_{0},q_{l_{0}}=0,\\
1\left(k_{l}=0\right), & l\neq l_{0},q_{l}\neq0,,l\notin\mathcal{C}_{m_{0}}\\
\overline{\Psi}\left(\lambda^{u}\rho_{l},\frac{q_{l}\lambda^{b}}{M}\right), & l\neq l_{0},q_{l}\neq0,l\in\mathcal{C}_{m_{0}},\\
\overline{\Psi}\left(\lambda^{u}\rho_{l},\lambda^{b}\right), & l\neq l_{0},q_{l}=0,
\end{cases}\label{eq:pmf-kl-conditional}
\end{align}
where $\mathcal{C}_{m_{0}}$ is the set of content files stored in
the cache of the BSs in $\Phi_{m_{0}}^{b}$,
\begin{equation}
\Psi\left(x,y\right)=\frac{3.5^{4.5}}{\Gamma\left(4.5\right)}\frac{x^{k-1}}{y^{k-1}\left(k-1\right)!}\frac{\Gamma\left(k+3.5\right)}{\left(\frac{x}{y}+3.5\right)^{k+3.5}},
\end{equation}
\begin{equation}
\overline{\Psi}\left(x,y\right)=\frac{3.5^{4.5}}{\Gamma\left(4.5\right)}\frac{x^{k}}{y^{k}k!}\frac{\Gamma\left(k+4.5\right)}{\left(\frac{x}{y}+3.5\right)^{k+4.5}}.
\end{equation}
\end{lemma}

The proof can be found in Appendix \ref{subsec:Probability-Mass-Function-K}. 

Using (\ref{eq:pmf-kl-conditional}), we calculate the expectation
of $K_{l}$ conditioned on $l_{0}$ and $m_{0}$, which is given by
\begin{align}
 & \mathbb{E}\left[K_{l}\big|L_{0}=l_{0},M_{0}=m_{0}\right]\nonumber \\
 & =\sum_{k_{l}=0}^{\infty}\Pr\left[K_{l}=k_{l}\big|L_{0}=l_{0},M_{0}=m_{0}\right]k_{l}\\
 & =\begin{cases}
1+\frac{9}{7}\frac{\lambda^{u}\rho_{l}}{\frac{q_{l}\lambda^{b}}{M}}, & l=l_{0},q_{l_{0}}\neq0,\\
1+\frac{9}{7}\frac{\lambda^{u}\rho_{l}}{\lambda^{b}}, & l=l_{0},q_{l_{0}}=0,\\
0, & l\neq l_{0},q_{l}\neq0,l\notin\mathcal{C}_{m_{0}}\\
\frac{9}{7}\frac{\lambda^{u}\rho_{l}}{\frac{q_{l}\lambda^{b}}{M}}, & l\neq l_{0},q_{l}\neq0,l\in\mathcal{C}_{m_{0}},\\
\frac{9}{7}\frac{\lambda^{u}\rho_{l}}{\lambda^{b}}, & l\neq l_{0},q_{l}=0.
\end{cases}
\end{align}
Then we calculate $\mathbb{E}\left[K_{l}\big|L_{0}=l_{0}\right]$
by removing the condition on $m_{0}$, given by
\begin{equation}
\mathbb{E}\left[K_{l}\big|L_{0}=l_{0}\right]=\begin{cases}
1+\frac{9}{7}\frac{\lambda^{u}\rho_{l}}{\lambda^{b}}, & l=l_{0},\\
\frac{9}{7}\frac{\lambda^{u}\rho_{l}}{\lambda^{b}}, & l\neq l_{0}.
\end{cases}
\end{equation}
Finally, we calculate $\mathbb{E}\left[K_{l}\right]$ by removing
the condition on $l_{0}$, given by
\begin{equation}
\mathbb{E}\left[K_{l}\right]=\rho_{l}+\frac{9}{7}\frac{\lambda^{u}\rho_{l}}{\lambda^{b}}.
\end{equation}

\subsubsection{Probability Mass Function of $\mathbf{K}$\label{subsec:Probability-Mass-Function-K}}

The probability mass function of $\mathbf{K}$ depends on the probability
density function of the size of the Voronoi cell of BS $B_{0}$ w.r.t.
content file $l\in\left\{ 1,\dots,L\right\} $. Denote $f_{Z}\left(z\right)$
as the probability density function of the size of the Voronoi cell
to which a randomly chosen user belongs, where $Z$ is a random variable
that denotes the size of the Voronoi cell normalized by the inverse
of the density of BSs. A widely used approximated form of this probability
density function is given by \cite{yu2013downlink}
\begin{equation}
f_{Z}\left(z\right)=\frac{3.5^{4.5}}{\Gamma\left(4.5\right)}z^{3.5}\exp\left(-3.5z\right).
\end{equation}

We first prove $\Pr\left[K_{l}=k_{l}\big|L_{0}=l_{0},M_{0}=m_{0}\right]$
in (\ref{eq:pmf-kl-conditional}):
\begin{itemize}
\item $l=l_{0}$ and $q_{l_{0}}\neq0$: The users requesting the $l$-th
content file form a homogeneous PPP with density $\lambda_{u}\rho_{l}$,
and the BSs that store the $l$-th content file form a homogeneous
PPP with density $\lambda_{b}q_{l}/M$. The probability mass function
of $K_{l}$ conditioned on $l_{0}$, $m_{0}$ and $z$ is given by
\begin{align}
 & \Pr\left[K_{l}=k_{l}\big|L_{0}=l_{0},M_{0}=m_{0},Z=z\right]\nonumber \\
 & =\frac{\left(\frac{\lambda^{u}\rho_{l}}{\lambda^{b}q_{l}/M}z\right)^{k_{l}}}{k_{l}!}e^{-\frac{\lambda^{u}\rho_{l}}{\lambda^{b}q_{l}/M}z}.
\end{align}
Then we calculate $\Pr\left[K_{l}=k_{l}\big|L_{0}=l_{0},M_{0}=m_{0}\right]$
by removing the condition on $z$, given by
\begin{align}
 & \Pr\left[K_{l}=k_{l}\big|L_{0}=l_{0},M_{0}=m_{0}\right]\nonumber \\
 & =\int_{0}^{\infty}\Pr\left[K_{l}=k_{l}-1\big|L_{0}=l_{0},M_{0}=m_{0},Z=z\right]\nonumber \\
 & \ \times f_{Z}\left(z\right)\mathrm{d}z\\
 & =\int_{0}^{\infty}\frac{\left(\frac{\lambda^{u}\rho_{l}}{\lambda^{b}q_{l}/M}z\right)^{k_{l}-1}}{\left(k_{l}-1\right)!}e^{-\frac{\lambda^{u}\rho_{l}}{\lambda^{b}q_{l}/M}z}f_{Z}\left(z\right)\mathrm{d}z\\
 & =\frac{3.5^{4.5}}{\Gamma\left(4.5\right)}\frac{\left(\lambda^{u}\rho_{l}\right)^{k_{l}-1}}{\left(\lambda^{b}q_{l}/M\right)^{k_{l}-1}\left(k_{l}-1\right)!}\nonumber \\
 & \ \times\int_{0}^{\infty}z^{k_{l}+2.5}\exp\left(-\left(\frac{\lambda^{u}\rho_{l}}{\lambda^{b}q_{l}/M}+3.5\right)z\right)\mathrm{d}z\\
 & =\Psi\left(\lambda^{u}\rho_{l},\frac{q_{l}\lambda^{b}}{M}\right).\label{eq:case1}
\end{align}
\item $l=l_{0}$ and $q_{l_{0}}=0$: The users requesting the $l$-th content
file form a homogeneous PPP with density $\lambda^{u}\rho_{l}$. All
BSs access to $l$-th content file via the backhaul, and they form
a homogeneous PPP with density $\lambda^{b}$. Similar to (\ref{eq:case1}),
we have
\begin{align}
 & \Pr\left[K_{l}=k_{l}\big|L_{0}=l_{0},M_{0}=m_{0}\right]\nonumber \\
= & \int_{0}^{\infty}\Pr\left[K_{l}=k_{l}-1\big|L_{0}=l_{0},M_{0}=m_{0},Z=z\right]\nonumber \\
 & \times f_{Z}\left(z\right)\mathrm{d}z\\
= & \Psi\left(\lambda^{u}\rho_{l},\lambda^{b}\right).
\end{align}
\item $l\neq l_{0}$, $q_{l_{0}}\neq0$ and $l\notin\mathcal{C}_{m_{0}}$:
Note the $q_{l_{0}}\neq0$ indicates that the $l$-th content file
is stored in some of the BSs. Meanwhile, $l\notin\mathcal{C}_{m_{0}}$
indicates the BSs in $\Phi_{m_{0}}^{b}$ do not cache the $l$-th
content file. As a result, the BSs in $\Phi_{m_{0}}^{b}$ do not serve
the users requesting the $l$-th content file, which means that $k_{l}=0$
is always satisfied. 
\item $l\neq l_{0}$, $q_{l_{0}}\neq0$ and $l\in\mathcal{C}_{m_{0}}$:
The users requesting the $l$-th content file form a homogeneous PPP
with density $\lambda^{u}\rho_{l}$, and the BSs that store the $l$-th
content file form a homogeneous PPP with density $\lambda^{b}q_{l}/M$.
The conditional probability mass function is given by
\begin{align}
 & \Pr\left[K_{l}=k_{l}\big|L_{0}=l_{0},M_{0}=m_{0}\right]\nonumber \\
= & \int_{0}^{\infty}\Pr\left[K_{l}=k_{l}\big|L_{0}=l_{0},M_{0}=m_{0},Z=z\right]\nonumber \\
 & \ \times f_{Z}\left(z\right)\mathrm{d}z\\
= & \int_{0}^{\infty}\frac{\left(\frac{\lambda^{u}\rho_{l}}{\lambda^{b}q_{l}/M}z\right)^{k_{l}}}{k_{l}!}e^{-\frac{\lambda_{u}\rho_{l}}{\lambda^{b}q_{l}/M}z}f_{Z}\left(z\right)\mathrm{d}z\\
= & \frac{3.5^{4.5}}{\Gamma\left(4.5\right)}\frac{\left(\lambda^{u}\rho_{l}\right)^{k_{l}}}{\left(\lambda^{b}q_{l}/M\right)^{k_{l}}k_{l}!}\nonumber \\
 & \times\int_{0}^{\infty}z^{k_{l}+3.5}\exp\left(-\left(\frac{\lambda^{u}\rho_{l}}{\lambda^{b}q_{l}/M}+3.5\right)z\right)\mathrm{d}z\\
= & \overline{\Psi}\left(\lambda^{u}\rho_{l},\frac{q_{l}\lambda^{b}}{M}\right).\label{eq:case4}
\end{align}
\item $l\neq l_{0}$, $q_{l_{0}}=0$: The users requesting the $l$-th content
file form a homogeneous PPP with density $\lambda^{u}\rho_{l}$. All
BSs access to $l$-th content file via the backhaul, and they form
a homogeneous PPP with density $\lambda^{b}$. Similar to (\ref{eq:case4}),
we have
\begin{align}
 & \Pr\left[K_{l}=k_{l}\big|L_{0}=l_{0},M_{0}=m_{0}\right]\nonumber \\
= & \int_{0}^{\infty}\Pr\left[K_{l}=k_{l}\big|L_{0}=l_{0},M_{0}=m_{0},Z=z\right]\nonumber \\
 & \times f_{Z}\left(z\right)\mathrm{d}z\\
= & \overline{\Psi}\left(\lambda^{u}\rho_{l},\lambda^{b}\right).
\end{align}
\end{itemize}
Finally, we calculate $\Pr\left[\mathbf{K}=\mathbf{k}\big|L_{0}=l_{0}\right]$
by removing the condition on $m_{0}$. For $q_{l_{0}}\neq0$, $m_{0}$
is selected from the BS groups $\mathcal{M}_{l_{0}}$ with equal probability.
For $q_{l_{0}}=0$, $m_{0}$ is selected from all the BS groups $\left\{ 0,\dots,M-1\right\} $
with equal probability. Hence, $\Pr\left[\mathbf{K}=\mathbf{k}\big|L_{0}=l_{0}\right]$
is given by (\ref{eq:pmf-k-1})\textendash (\ref{eq:pmf-k-2}) by
removing the condition on $m_{0}$ in $\Pr\left[\mathbf{K}=\mathbf{k}\big|L_{0}=l_{0},M_{0}=m_{0}\right]$.

\subsection{Expression of Successful Transmission Probability $p$ \label{subsec:Accurate-Expression-of-p}}

The BSs loading $\mathbf{K}$ and SIR are correlated, since BSs with
a larger cell size have higher loading and lower SIR \cite{singh2014joint}.
However, the exact relationship between $\mathbf{K}$ and SIR is very
complex and is still not known. For tractability of the analysis,
as in \cite{singh2014joint}, the dependence is ignored. Hence, the
successful transmission probability conditioned on the $l_{0}$-th
content file requested by $u_{0}$ and distance $d$ is given by
\begin{align}
p_{l_{0},d}= & \sum_{\mathbf{k}\in\mathbb{N}^{L}}\Pr\left[\mathbf{K}=\mathbf{k}\big|L_{0}=l_{0}\right]\nonumber \\
 & \times\Pr\left[C\geq\tau\big|\mathbf{K}=\mathbf{k},L_{0}=l_{0},S=1\right]\nonumber \\
 & \times\Pr\left[S=1\big|\mathbf{K}=\mathbf{k},L_{0}=l_{0}\right],
\end{align}
where $\Pr\left[C\geq\tau\big|\mathbf{K}=\mathbf{k},L_{0}=l_{0},S=1\right]$
follows (\ref{eq:phy-outage}), and $\Pr\left[S=1\big|\mathbf{K}=\mathbf{k},L_{0}=l_{0}\right]$
follows (\ref{eq:backhaul-outage}). Then we calculate $p_{l_{0}}$
by removing the condition on $d$:
\begin{equation}
p_{l_{0}}=\int_{0}^{\infty}p_{l_{0},d}\left(d\right)f_{D_{0,0},l_{0}}\left(d\right)\mathrm{d}d.
\end{equation}
Finally, by the total probability theorem, the average successful
transmission probability is given by
\begin{equation}
p=\sum_{l_{0}=1}^{L}p_{l_{0}}\rho_{l_{0}}.
\end{equation}

Note that the above expression of $p$ is complicated, since the expression
of the probability mass function of $\mathbf{k}$ is complicated.
As a result, it is hard to find the optimal number of subbands $M$
and cache storage allocation vector $\mathbf{q}$ that maximize the
average successful transmission probability $p$.

\subsection{Proof of Theorem \ref{thm:Optimal-solution-sub} \label{subsec:Proof-of-Theorem-optimal-solution}}

By removing constraint (\ref{eq:ql-order}) in Problem $\tilde{\mathcal{P}}\left(M,L^{\prime}\right)$,
we obtain the following problem:
\begin{align}
\check{\mathcal{P}}\left(M,L^{\prime}\right):\min_{q_{l}} & \sum_{l=1}^{L^{\prime}}\rho_{l}\frac{\beta\left(M,\tilde{g}_{0}\right)}{q_{l_{0}}+\beta\left(M,\tilde{g}_{0}\right)}\\
\mathrm{s.t.}\  & \sum_{l=1}^{L^{\prime}}q_{l}=MB_{C},\label{eq:L-1}\\
 & q_{l}\geq1,\forall l\in\left\{ 1,\dots,L^{\prime}\right\} ,\label{eq:L-2}\\
 & q_{l}\leq M,\forall l\in\left\{ 1,\dots,L^{\prime}\right\} .\label{eq:L-3}
\end{align}
Denote $\check{\mathbf{q}}^{\star}=\left[\check{q}_{1}^{\star},\dots,\check{q}_{L^{\prime}}^{\star}\right]$
as the optimal solution of $\check{\mathcal{P}}\left(M,L^{\prime}\right)$.
We will show later that an optimal solution of $\check{\mathcal{P}}\left(M,L^{\prime}\right)$
is also an optimal solution of $\tilde{\mathcal{P}}\left(M,L^{\prime}\right)$.
Problem $\check{\mathcal{P}}\left(M,L^{\prime}\right)$ is a convex
optimization problem. Introducing a Lagrange multiplier $\lambda^{\star}$
for equality constraint (\ref{eq:L-1}), multipliers $\nu_{l}^{\star}$
for constraint (\ref{eq:L-2}), and multipliers $\omega_{l}^{\star}$
for constraint (\ref{eq:L-3}), we obtain the KKT conditions:
\begin{equation}
\sum_{l=1}^{L^{\prime}}\check{q}_{l}^{\star}=MB_{C},\check{q}_{l}^{\star}\geq1,\check{q}_{l}^{\star}\leq M.
\end{equation}
\begin{equation}
\nu^{\star}\geq0,\omega^{\star}\geq0.
\end{equation}
\begin{equation}
\nu_{l}^{\star}\left(1-\check{q}_{l}^{\star}\right)=0,\omega_{l}^{\star}\left(\check{q}_{l}^{\star}-M\right)=0.
\end{equation}
\begin{equation}
-\frac{\rho_{l}}{\left(\check{q}_{l}^{\star}+\beta\left(M,\tilde{g}_{0}\right)\right)^{2}}+\lambda^{\star}-\nu_{l}^{\star}+\omega_{l}^{\star}=0,\forall l.
\end{equation}
By eliminating $\nu$, we have 
\begin{equation}
\sum_{l=1}^{L^{\prime}}\check{q}_{l}^{\star}=MB_{C},\check{q}_{l}^{\star}\geq1,\check{q}_{l}^{\star}\leq M.
\end{equation}
\begin{equation}
\omega^{\star}\geq0.
\end{equation}
\begin{equation}
\left(\lambda^{\star}-\frac{\rho_{l}}{\left(\check{q}_{l}^{\star}+\beta\left(M,\tilde{g}_{0}\right)\right)^{2}}+\omega_{l}^{\star}\right)\left(1-\check{q}_{l}^{\star}\right)=0,
\end{equation}
\begin{equation}
\omega_{l}^{\star}\left(\check{q}_{l}^{\star}-M\right)=0.
\end{equation}
\begin{equation}
\lambda^{\star}\geq\frac{\rho_{l}}{\left(\check{q}_{l}^{\star}+\beta\left(M,\tilde{g}_{0}\right)\right)^{2}}-\omega_{l}^{\star},\forall l.
\end{equation}

\begin{itemize}
\item If $\lambda^{\star}<\frac{\rho_{l}}{\left(1+\beta\left(M,\tilde{g}_{0}\right)\right)^{2}}-\omega_{l}^{\star}$,
the last condition can only hold if $\check{q}_{l}^{\star}>1$, which
by the third condition implies that $\lambda^{\star}=\frac{\rho_{l}}{\left(\check{q}_{l}^{\star}+\beta\left(M,\tilde{g}_{0}\right)\right)^{2}}-\omega_{l}^{\star}$,
i.e., $\check{q}_{l}^{\star}=\sqrt{\frac{\rho_{l}}{\lambda^{\star}+\omega_{l}^{\star}}}-\beta\left(M,\tilde{g}_{0}\right)$. 
\begin{itemize}
\item If $\sqrt{\frac{\rho_{l}}{\lambda^{\star}+\omega_{l}^{\star}}}-\beta\left(M,\tilde{g}_{0}\right)\geq M$,
since $\check{q}_{l}^{\star}>M$ is impossible, we have $\check{q}_{l}^{\star}=M$
and $\frac{\rho_{l}}{\lambda^{\star}+\omega_{l}^{\star}}=\left(M+\beta\left(M,\tilde{g}_{0}\right)\right)^{2}$,
i.e., $\omega_{l}^{\star}=\frac{\rho_{l}}{\left(M+\beta\left(M,\tilde{g}_{0}\right)\right)^{2}}-\lambda^{\star}$.
Since $\lambda^{\star}\leq\frac{\rho_{l}}{\left(M+\beta\left(M,\tilde{g}_{0}\right)\right)^{2}}$,
$\omega_{l}^{\star}\geq0$ is feasible.
\item If $\sqrt{\frac{\rho_{l}}{\lambda^{\star}+\omega_{l}^{\star}}}-\beta\left(M,\tilde{g}_{0}\right)<M$,
then $\omega_{l}^{\star}=0$. We then have $\sqrt{\frac{\rho_{l}}{\lambda^{\star}}}-\beta\left(M,\tilde{g}_{0}\right)<M$,
which indicates $\lambda^{\star}>\frac{\rho_{l}}{\left(M+\beta\left(M,\tilde{g}_{0}\right)\right)^{2}}$,
$\check{q}_{l}^{\star}=\sqrt{\frac{\rho_{l}}{\lambda^{\star}}}-\beta\left(M,\tilde{g}_{0}\right)$.
\end{itemize}
\item If $\lambda^{\star}\geq\frac{\rho_{l}}{\left(1+\beta\left(M,\tilde{g}_{0}\right)\right)^{2}}-\omega_{l}^{\star}$,
then $\check{q}_{l}^{\star}>1$ is impossible, because it would imply
$\lambda^{\star}\geq\frac{\rho_{l}}{\left(1+\beta\left(M,\tilde{g}_{0}\right)\right)^{2}}-\omega_{l}^{\star}>\frac{\rho_{l}}{\left(\hat{q}_{l}^{\star}+\beta\left(M,\tilde{g}_{0}\right)\right)^{2}}-\omega_{l}^{\star}$,
which violates the complementary slackness condition. Therefore, we
have $\check{q}_{l}^{\star}=1$.
\end{itemize}
In summary, the optimal solution is given by
\begin{align}
 & \check{q}_{l}^{\star}\nonumber \\
 & =\begin{cases}
M, & \lambda^{\star}\leq\frac{\rho_{l}}{\left(M+\beta\left(M,\tilde{g}_{0}\right)\right)^{2}},\\
\sqrt{\frac{a_{l}}{\lambda^{\star}}}-\beta\left(M,\tilde{g}_{0}\right), & \frac{\rho_{l}}{\left(M+\beta\left(M,\tilde{g}_{0}\right)\right)^{2}}<\lambda^{\star}<\frac{\rho_{l}}{\left(1+\beta\left(M,\tilde{g}_{0}\right)\right)^{2}}\\
1, & \lambda^{\star}\geq\frac{a_{l}}{\left(1+\beta\left(M,\tilde{g}_{0}\right)\right)^{2}},
\end{cases},
\end{align}
which is equivalent to
\begin{equation}
\check{q}_{l}^{\star}=\min\left\{ M,\max\left\{ 1,\sqrt{\rho_{l}/\lambda^{\star}}-\beta\left(M,\tilde{g}_{0}\right)\right\} \right\} .
\end{equation}
Substituting this expression for $\check{q}_{l}^{\star}$ into $\sum_{l=1}^{L^{\prime}}\check{q}_{l}^{\star}=MB_{C}$,
we obtain

\begin{equation}
\sum_{l=1}^{L^{\prime}}\min\left\{ M,\max\left\{ 1,\sqrt{\rho_{l}/\lambda^{\star}}-\beta\left(M,\tilde{g}_{0}\right)\right\} \right\} =MB_{C}.
\end{equation}
Note that since $\rho_{l}\geq\rho_{l+1}$ for $l=1,\dots,L^{\prime}-1$,
$\check{q}_{l}^{\star}$ satisfies $\check{q}_{l}^{\star}\geq\check{q}_{l+1}^{\star}$
for $l=1,\dots,L^{\prime}-1$. As a result, $\check{\mathbf{q}}$
is also an optimal solution of $\tilde{\mathcal{P}}\left(M,L^{\prime}\right)$. 

\bibliographystyle{IEEEtran}


\begin{IEEEbiography}[{\includegraphics[width=1in,height=1.25in,clip,keepaspectratio]{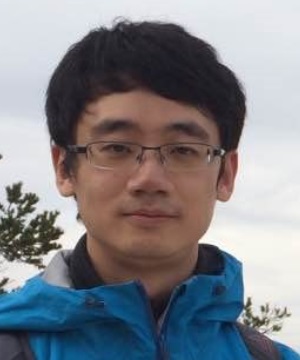}}]{Wei Han (S'12)}received the B.S. from Tsinghua University (2007-2011) and Ph.D. from the Hong Kong University of Science and Technology (HKUST) (2012-2018). He is currently  a researcher with Future Network Theory Lab, 2012 Labs, Huawei Tech. Investment Co., Ltd.. His research interests include wireless caching and compressive sensing.  \end{IEEEbiography}

\begin{IEEEbiography}[{\includegraphics[width=1in,height=1.25in,clip,keepaspectratio]{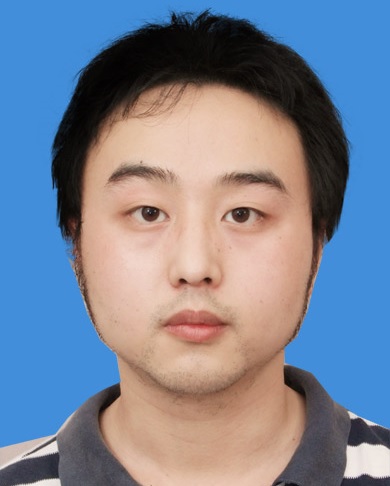}}]{An Liu (S'07-M'09-SM'17)}received the Ph.D. and the B.S. degree in Electrical Engineering from Peking University, China, in 2011 and 2004 respectively. From 2008 to 2010, he was a visiting scholar at the Department of ECEE, University of Colorado at Boulder. He has been a Postdoctoral Research Fellow in 2011-2013, Visiting Assistant Professor in 2014, and Research Assistant Professor in 2015-2017, with the Department of ECE, HKUST. He is currently a Distinguished Research Fellow with the College of Information Science and Electronic Engineering, Zhejiang University. His research interests include wireless communications, stochastic optimization and compressive sensing.  \end{IEEEbiography}

\begin{IEEEbiography}[{\includegraphics[width=1in,height=1.25in,clip,keepaspectratio]{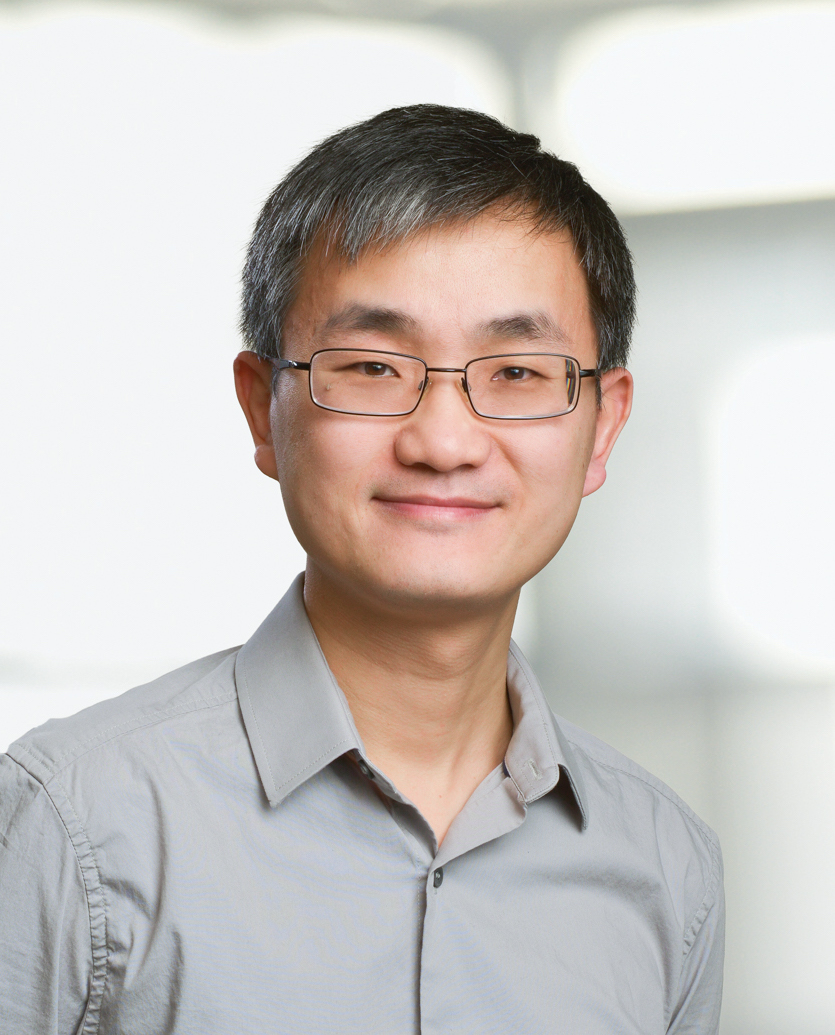}}]{Wei Yu (S'97-M'02-SM'08-F'14)} received the B.A.Sc. degree in Computer Engineering and Mathematics from the University of Waterloo, Waterloo, Ontario, Canada in 1997 and M.S. and Ph.D. degrees in Electrical Engineering from Stanford University, Stanford, CA, in 1998 and 2002, respectively. Since 2002, he has been with the Electrical and Computer Engineering Department at the University of Toronto, Toronto, Ontario, Canada, where he is now Professor and holds a Canada Research Chair (Tier 1) in Information Theory and Wireless Communications. His main research interests include information theory, optimization, wireless communications and broadband access networks.

Prof. Wei Yu currently serves on the IEEE Information Theory Society Board of Governors (2015-20). He was an IEEE Communications Society Distinguished Lecturer (2015-16). He is currently an Area Editor for the IEEE Transactions on Wireless Communications (2017-20). He served as an Associate Editor for IEEE Transactions on Information Theory (2010-2013), as an Editor for IEEE Transactions on Communications (2009-2011), and as an Editor for IEEE Transactions on Wireless Communications (2004-2007). He is currently the Chair of the Signal Processing for Communications and Networking Technical Committee of the IEEE Signal Processing Society (2017-18) and served as a member in 2008-2013. Prof. Wei Yu received the Steacie Memorial Fellowship in 2015, the IEEE Signal Processing Society Best Paper Award in 2017 and 2008, an Journal of Communications and Networks Best Paper Award in 2017, an IEEE Communications Society Best Tutorial Paper Award in 2015, an IEEE ICC Best Paper Award in 2013, the McCharles Prize for Early Career Research Distinction in 2008, the Early Career Teaching Award from the Faculty of Applied Science and Engineering, University of Toronto in 2007, and an Early Researcher Award from Ontario in 2006. Prof. Wei Yu is a Fellow of the Canadian Academy of Engineering, and a member of the College of New Scholars, Artists and Scientists of the Royal Society of Canada. He is recognized as a Highly Cited Researcher.   \end{IEEEbiography}

\begin{IEEEbiography}[{\includegraphics[width=1in,height=1.25in,clip,keepaspectratio]{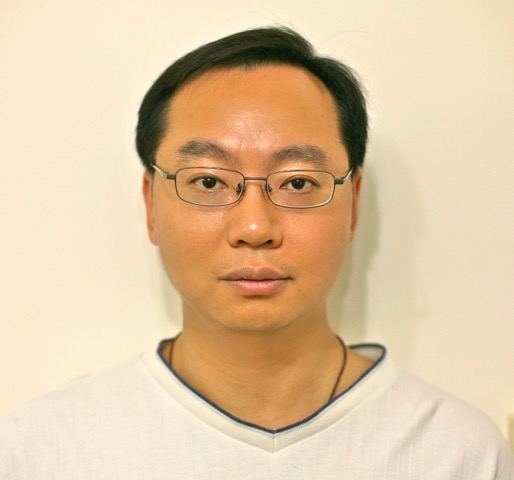}}]{Vincent K. N. Lau (SM'04-F'12)}obtained B.Eng (Distinction 1st Hons) from the University of Hong Kong (1989-1992) and Ph.D. from the Cambridge University (1995-1997). He joined Bell Labs from 1997-2004 and the Department of ECE, Hong Kong University of Science and Technology (HKUST) in 2004. He is currently a Chair Professor and the Founding Director of Huawei-HKUST Joint Innovation Lab at HKUST. His current research focus includes robust and delay-optimal cross layer optimization for MIMO/OFDM wireless systems, interference mitigation techniques for wireless networks, massive MIMO, M2M and network control systems.  \end{IEEEbiography}
\end{document}